\documentstyle[preprint,aps,epsfig]{revtex}
\tightenlines
\begin{document}
\draft
\preprint{\vbox{
\hbox{IFT-P.082/98}
\hbox{November 1998}
}}
\title{Soft superweak CP violation in a 331 model} 
\author{ J. C. Montero~\footnote{E-mail:montero@ift.unesp.br},
V. Pleitez~\footnote{E-mail:vicente@ift.unesp.br} and 
O. Ravinez~\footnote{Present address: 
Universidad Nacional de Ingeneria (UNI), Facultad de Ciencias
Av. Tupac Amaru S/N apartado 31139, Lima-Peru. E-mail:pereyra@fc-uni.edu.pe} }  
\address{
Instituto de F\'\i sica Te\'orica\\
Universidade Estadual Paulista\\
Rua Pamplona 145\\ 
01405-900--S\~ao Paulo, SP\\Brazil} 
\date{\today}
\maketitle
\begin{abstract}
We show that it is possible to implement soft superweak $CP$ violation in the 
context of a 331 model with only three triplets. All $CP$ violation effects 
come from the exchange of singly and doubly charged scalars. 
We consider the implication of this mechanism in the quark and lepton sectors. 
In particular it is shown that in this model, as in most of those which 
incorporate the scalar mediated $CP$ violation, it is possible to have large 
electric dipole moments for the muon and the tau lepton while keeping small 
those of the electron and the neutron. The CKM mixing matrix is real up to the two 
loop level.
\end{abstract}
\section{Introduction}
\label{sec:intro}
The origin and the smallness of $CP$ violation are still open questions.
In the context of the electroweak standard model~\cite{sm} the only source of
$CP$ violation, in the quark sector, is the surviving phase in the charged 
current coupled to the vector boson $W^{\pm}$~\cite{km}. This is called 
explicit or hard $CP$ violation. 
On the other hand there is no $CP$ violation in the lepton sector at lower
orders since neutrinos are massless. 

Although this is an interesting feature of the model it leaves open the 
question of why $CP$ is so feebly violated. 
It is well know that $CP$ is softly broken if it occurs through a dimensional 
coupling in the bare lagrangian and/or spontaneously. 
The possibility that $CP$ nonconservation arise exclusively through Higgs 
boson exchanges is a rather old one. It can be implemented in a
spontaneous way as it was proposed by Lee~\cite{tdlee} or, explicitly in
the parameters of the scalar potential, as proposed by Weinberg~\cite{sw1,dono}.
Of course, both possibilities can be mixed. 
Since the works of Lee and Weinberg it has been known that in 
renormalizable gauge theories the violation of $CP$ has the right strength
if it occurs through the exchange of a Higgs boson of mass $M_H$, 
{\it i. e.}, it is proportional to $G_Fm^2_f/M^2_H$, where $m_f$ is the fermion 
mass. Since then, there have been many realizations of that mechanism in 
extensions of the electroweak standard model.
Even if we insist that the $CP$ violation arises solely through the Higgs 
exchange we have several possibilities in the standard electroweak model with 
several doublets~\cite{lw}.

Some years ago it was proposed a model based on the
$SU(3)_C\otimes SU(3)_L\otimes U(1)_N$ gauge symmetry with exotic charged 
leptons~\cite{pt}. In this model since the scalar multiplets transform
in a different way one from another the scalar potential is constrained
in such a way that even with three triplets there is not spontaneous $CP$ 
violations~\cite{laplata1}. Here we will show that it is possible to have
soft $CP$ violation because the coupling constant of the trilinear term 
in the scalar potential is complex and the vacuum expectation value (VEV) of 
the three triplets are also complex. Hence, $CP$ is a symmetry in the full bare 
lagrangian but in the trilinear term in the potential, in particular all 
Yukawa couplings are real at tree level. We will show that the model is 
a realization of the pure superweak $CP$ violation~\cite{lw2} in the sense that 
all flavor changing phenomena other than $CP$ violation are
accurately described by the real Cabbibo-Kobayashi-Maskawa (CKM) 
matrix since $CP$ violation is restricted to one operator of dimension 
three~\cite{rb}.

If the condition that except in the trilinear term $CP$ is a 
symmetry of the lagrangian is assumed, we have
verified that it is possible to choose the physical phases in such a way that 
the only places of $CP$ violation in the lagrangian density
are those related with the singly and doubly charged scalars which are present 
in the model. One interesting feature of the model is that the electric dipole 
moment of the muon and tau lepton can be of the order of magnitude of the 
respective present experimental bounds. In the lepton tau case it means that 
this quantity can be studied in a tau-charm factory.

This work is organized as follows. In Sec.~\ref{sec:m1} we review the Higgs 
sector of the model. We consider in particular the minimization of the scalar 
potential and we also give the definitions of the Goldstone and the physical 
scalar fields of the several charged sectors. In Sec.~\ref{sec:yuka} we study 
the Yukawa interactions showing how the VEV's phases can be 
absorbed in the fermion field in some places of the lagrangian density and
that they only survive in the sector involving both simply and doubly charged
scalar fields. In Sec.~\ref{sec:gq} we show that there is not $CP$ violation
in the vector boson-fermion interactions. The phenomenology of the model is
considered in Sec.~\ref{sec:feno} while our conclusions are in the last section.

\section{A model with three scalar triplets}
\label{sec:m1}
As we said before, here we will consider a model with 
$SU(3)_C\otimes SU(3)_L\otimes U(1)_N$ symmetry with both exotic quarks and
charged leptons~\cite{pt}.
In this model in order to give mass to all fermions it is necessary to 
introduce three scalar triplets. They transform as
\begin{equation}
{\bf \chi }=\left( 
\begin{array}{c}
\chi ^{-} \\ 
\chi ^{--} \\ 
\chi ^0
\end{array}
\right) \sim \left( {\bf 3},{\bf -1}\right),  \qquad {\bf \rho }=\left( 
\begin{array}{c}
\rho ^{+} \\ 
\rho ^0 \\ 
\rho ^{++}
\end{array}
\right) \sim \left( {\bf 3},{\bf 1}\right), \qquad 
{\bf \eta }=\left( 
\begin{array}{c}
\eta ^0 \\ 
\eta _1^{-} \\ 
\eta _2^{+}
\end{array}
\right) \sim \left( {\bf 3},{\bf 0}\right) .
\label{e1}
\end{equation}

The more general scalar potential invariant under the gauge symmetry is 
\begin{eqnarray}
V(\chi,\eta,\rho)&=&\mu _1^2\chi^{\dagger } \chi +\mu _2^2\eta^\dagger\eta +
\mu _3^2\rho^\dagger\rho+(\alpha \epsilon_{ijk}\chi_i \rho_j \eta_k +H.c.) 
\nonumber \\ &&
+a_1\left(\chi^\dagger\chi \right)^2+
a_2\left(\eta^\dagger\eta\right)^2+a_3\left( \rho^\dagger \rho \right)^2+
a_4\left(\chi^\dagger\chi\right) \left(\eta^\dagger\eta\right) +
a_5\left(\chi^\dagger\chi\right) \left(\rho^\dagger\rho \right) 
 \nonumber \\
&&+a_6\left(\rho^\dagger\rho\right) \left(\eta^\dagger\eta\right) +
a_7\left(\chi^\dagger\eta \right) \left(\eta^\dagger\chi\right) +
a_8\left(\chi^\dagger\rho\right) \left(\rho^\dagger\chi\right) +
a_9\left(\rho^\dagger\eta\right) \left(\eta^\dagger\rho\right)  \nonumber \\
&&+\left[a_{10}\left(\chi^\dagger\eta\right) \left( 
\rho^\dagger\eta \right) +H.c.\right].
\label{potential}
\end{eqnarray}
All terms of this potential but the $a_{10}$ term conserve the total lepton 
number $L$ (or $L+B$, where $B$ is the baryonic number).
The minimum of the potential must be studied after the shifting of the
neutral components of the three scalar multiplets in Eq.~(\ref{e1}). Hence, we
redefine the neutral components as follows:
\begin{equation}
\eta ^0\to \frac{v_\eta }{\sqrt{2}}\left( 1+
\frac{ X^0_\eta +iI^0_\eta}{\vert v_\eta\vert }\right) ,\;\;
\rho ^0\to \frac{v_\rho }{\sqrt{2}}\left( 1+
\frac{ X^0_\rho +iI^0_\rho}{\vert v_\rho\vert}\right), 
\;\;
\chi^0\to \frac{ v_\chi}{\sqrt{2}}\,
\left( 1+\frac{X^0_\chi+iI^0_\chi}{\vert v_\chi\vert}\right),
\label{vev}
\end{equation}
where $v_a=\vert v_a\vert\,e^{i\theta_a}$ with $a=\eta,\rho,\chi$.

The condition that the first derivative of the potential in 
Eq.~(\ref{potential}) is zero ({\it i.e}, no linear terms in all neutral fields
must survive) implies the following constraint equations:
\begin{mathletters}
\label{vin}
\begin{equation}
\frac{a_4}2\left| v_\eta \right| ^2\left| v_\chi\right| +\frac{a_5}2\left|
v_\rho \right| ^2\left| v_\chi\right| +\mu _1^2\left| v_\chi\right| +a_1\left|
v_\chi\right| ^3+\frac 1{\sqrt{2}\left| v_\chi\right| }{\rm Re}\left( \alpha
v_\chi v_\rho v_\eta \right) =0, 
\label{vin1}
\end{equation}

\begin{equation}
a_2\left| v_\eta \right| ^3+\frac{a_4}2\left| v_\chi\right| ^2\left| v_\eta
\right| +\frac{a_6}2\left| v_\rho \right| ^2\left| v_\eta \right| +\mu
_2^2\left| v_\eta \right| +\frac 1{\sqrt{2}\left| v_\eta \right| }{\rm Re}%
\left( \alpha v_\chi v_\rho v_\eta \right) =0, 
\label{vin2}
\end{equation}

\begin{equation}
\frac{a_6}2\left| v_\eta \right| ^2\left| v_\rho \right| +\frac{a_5}2\left|
v_\rho \right| \left| v_\chi\right| ^2+\mu _3^2\left| v_\rho \right| +a_3\left|
v_\rho \right| ^3+\frac 1{\sqrt{2}\left| v_\rho \right| }{\rm Re}\left(
\alpha v_\chi v_\rho v_\eta \right) =0, 
\label{vin3}
\end{equation}
and, finally
\begin{equation}
{\rm Im}\left( \alpha v_\chi v_\rho v_\eta \right) =0.  
\label{vin4}
\end{equation}
\end{mathletters}

Before considering how many physical phases will survive in 
the potential (this must be done by taking at the same time the Yukawa 
interactions) we will study all scalar mass eigenstates in the model. 
Recall that we have to verify where the VEV's phases appear in the several
sectors of the lagrangian that is, in vertices and in mixing matrices. Hence, 
we will firstly write down explicitly the mass and mixing matrices of each 
charged sector of the model. Two of the phases in Eqs.~(\ref{vev}) can be 
transformed away with a $SU(3)$ transformation; whenever we do that
we will mention it explicitly and we will choose $\theta_\eta=\theta_\rho=0$.

\subsection{Doubly charged scalars}
\label{subsec:2cs}
In this sector we have the following 
mass matrix in the $(\rho^{++},\;\chi^{++})^T$ basis
\begin{equation}
\left( \begin{array}{cc}
-\frac{a_8}2\left| v_\chi\right| ^2+\frac A{\sqrt{2}\left| v_\rho
\right| ^2}\; &\;
\frac{\alpha _1^{*}v_\eta^{*}}{\sqrt{2}}-\frac{a_8}2v_\rho v_\chi \\
\;&\; -\frac{a_8}2\left| v_\rho \right| ^2+\frac A{\sqrt{2}\left|
v_\chi\right| ^2}
\end{array}\right)
\label{dcm1}
\end{equation}
where we have defined $A\equiv{\rm Re}\left( \alpha v_\chi v_\eta v_\rho 
\right)$.

As expected, we have a doubly charged Goldstone boson $G^{++}$ and a physical 
doubly charged scalar $Y^{++}$ 
\begin{equation}
\left( 
\begin{array}{c}
\rho^{++} \\ \chi^{++}
\end{array}
\right)=\frac{1}{(\vert v_\rho\vert^2+\vert v_\chi\vert^2)^{\frac{1}{2}}}
\left(\begin{array}{cc}
\vert v_\rho\vert & -\vert v_\chi\vert\,e^{-i\theta_\chi} 
\\ \vert v_\chi\vert\,e^{i\theta_\chi}  & \vert v_\rho\vert 
\end{array}
\right)
\left( 
\begin{array}{c}
G^{++} \\ Y^{++}
\end{array}
\right),
\label{dcf}
\end{equation}
with the mass square of the $Y^{++}$ field given by
\begin{equation}
M^2_{Y^{++}}=\frac A{\sqrt{2}}\left( \frac 1{\left| v_\chi\right| ^2}+
\frac1{\left| v_\rho \right| ^2}\right)
-\frac{a_8}2\left( \left| v_\chi\right| ^2+\left| v_\rho \right|^2\right) 
\label{m++}
\end{equation}
Notice that this mass is proportional to $\vert v_\chi\vert^2$ which is the
VEV that is in control of the $SU(3)_L$ symmetry and so it is the largest mass
scale of the model. It means that the doubly charged scalar must be a heavy 
scalar.

\subsection{Singly charged scalars}
\label{subsec:1cs}
Next, for the simply charged scalar fields we have, in the $(\eta^+_1,\,\rho^+,
\,\eta^+_2,\,\chi^+)^T$ basis,
\begin{equation}
\left(
\begin{array}{cccc}
-\frac{a_9}2\left| v_\rho \right|^2+\frac A{\sqrt{2}\left| v_\eta\right|^2} 
\;&\;
-\frac{a_9}2v_\rho ^{*}v_\eta ^{*}+\frac{\alpha v_\chi}{\sqrt{2}}\;&\;
-\frac{a_{10}}2v_\rho ^{*}v_\chi^{*}& 
-\frac{a_{10}}2v_\rho ^{*}v_\eta\\
 & -\frac{a_9}2\left| v_\eta \right| ^2+\frac A{\sqrt{2}\left| v_\rho
\right| ^2} & -\frac{a_{10}}2v_\chi^{*}v_\eta &
-\frac{a_{10}}2v_\eta ^2 \\
& & -\frac{a_7}2\left| v_\chi\right| ^2+\frac A{\sqrt{2}\left| v_\eta
\right|^2} & -\frac{a_7}2v_\chi v_\eta +\frac{\alpha^{*}v_\rho^{*}}
{\sqrt{2}}\\
& & & -\frac{a_7}2\left| v_\eta \right| ^2+\frac A{\sqrt{2}\left|
v_\chi\right| ^2}
\end{array}
\right).
\label{scm}
\end{equation}

Notice that if $a_{10}=0$, ${\eta}_1^+$ and $\rho^+$ decouple from
${\eta}^+_2$ and $\chi^+$. Hence, the mass matrix in Eq.~(\ref{scm}) is 
reduced to two $2\times 2$ mass matrices and we have two Goldstone bosons 
$G^+_1$  and $G^+_2$; and two massive fields $Y^+_1$ and $Y^+_2$,
\begin{equation}
\left(
\begin{array}{c}
{\eta}_1^+ \\ \rho^+
\end{array} \right)=
\frac 1{(\vert v_\eta\vert^2+\vert v_\rho\vert^2)^{\frac{1}{2}}}
\left(
\begin{array}{cc}
-\vert v_\eta\vert & \vert v_\rho\vert \\
\vert v_\rho\vert & \vert v_\eta\vert
\end{array}\right)
\left(
\begin{array}{c}
G^+_1\\ Y_1^+
\end{array}
\right),
\label{scf1}
\end{equation}
with 
\begin{equation}
m^2_{Y^+_1}=\frac{A}{\sqrt2}\left( \frac{1}{\vert v_\eta\vert^2}+\frac{1}
{\vert v_\rho\vert^2}\right)-\frac{a_9}{2}\left( \vert v_\rho\vert^2+
\vert v_\eta\vert^2\right),
\label{y1}
\end{equation}
and
\begin{equation}
\left(
\begin{array}{c}
{\eta}_2^+ \\ \chi^+
\end{array} \right)=
\frac 1{(\vert v_\eta\vert^2+\vert v_\chi\vert^2)^{\frac{1}{2}}}
\left(
\begin{array}{cc}
-\vert v_\eta\vert & \vert v_\chi\vert \\
-\vert v_\chi\vert & \vert v_\eta\vert
\end{array}\right)
\left(
\begin{array}{c}
G^+_2\\ Y^+_2
\end{array}
\right),
\label{scf2}
\end{equation}
with 
\begin{equation}
m^2_{Y^+_2}=\frac{A}{\sqrt2}\left( \frac{1}{\vert v_\eta\vert^2}+\frac{1}
{\vert v_\chi\vert^2}\right)-\frac{a_7}{2}\left( \vert v_\chi\vert^2+
\vert v_\eta\vert^2\right).
\label{y2}
\end{equation}

There are 5 phases in the matrix in Eq.~(\ref{scm}), two of them can be 
transformed away with a $SU(3)$ transformation (here when we do that
we will chose $\theta_\eta=\theta_\rho=0$). The other three phases can be
absorbed by redefining three scalar fields. However these phases will
appear in the Yukawa interactions or in the scalar potential in terms
like $a_{10}\chi^{+}\eta^0\rho^-\eta^0$. Hence, there is $CP$ violation
in the propagators of the singly charged scalars if $a_{10}\not=0$.
Since we want $CP$ softly broken we will consider here $a_{10}=0$ with
the singly charged scalar given by the expressions above.

\subsection{Neutral scalars}
\label{subsec:os}
Finally, in the neutral sector we have $CP$ even fields, denoted here by  
$H^0_i$, and $CP$ odd fields, $(G^0_{1,2},h^0)\equiv h_i^0$.  

In the $CP$ odd sector we have
\begin{equation}
\left(
\begin{array}{ccc}
\frac A{\sqrt{2}\left| v_\eta \right| ^2}&
 \frac A{\sqrt{2}\left| v_\eta \right| \left| v_\rho \right| }
 & \frac A{\sqrt{2}\left| v_\eta \right| \left| v_\chi\right| }
 \\
 & \frac A{\sqrt{2}\left| v_\rho \right| ^2}&
\frac A{\sqrt{2}\left| v_\chi\right| \left| v_\rho \right| } \\
& & \frac A{\sqrt{2}\left| v_\chi\right| ^2}
 \end{array}\right)
\label{nfm}
\end{equation}
in the $(I_\eta,I_\rho,I_\chi)^T$ basis. 
There are in fact two neutral Goldstone bosons $G^0_1$ and $G^0_2$ (as it must 
be since we have two massive neutral vector bosons $Z$ and $Z'$) and a physical
$CP$ odd scalar field $h^0$. Explicitly we have
\begin{equation}
\left(\begin{array}{c}
I^0_\eta  \\ I^0_\rho  \\ I^0_\chi\end{array}\right)=
\left(
\begin{array}{ccc}
-\frac{\vert v_\eta\vert}{N_1} & -\frac{\vert v_\rho\vert^2\vert v_\eta\vert}
{N_2} & \frac{\vert v_\rho\vert \vert v_\chi\vert}{N_3}\\
-\frac{\vert v_\rho\vert}{N_1} & 
-\frac{\vert v_\eta\vert^2\vert v_\rho\vert}{N_2}&
\frac{\vert v_\eta\vert \vert v_\chi\vert}{N_3}\\
0& \frac{\vert v_\chi\vert N_1^2}{N_2}& 
\frac{\vert v_\eta\vert \vert v_\rho\vert}{N_3}
\end{array}
\right)
\left(\begin{array}{c}
G^0_1\\ G^0_2 \\ h^0
\end{array}\right),
\label{nsi}
\end{equation}
where 
\begin{equation}
N_1=(\vert v_\eta\vert^2+\vert v_\rho\vert^2)^{\frac{1}{2}},\; 
N_2=(\vert v_\eta\vert^2\vert v_\rho\vert^4+
\vert v_\eta\vert^4\vert v_\rho\vert^2 +\vert v_\chi\vert^2N_1^4)^{\frac{1}{2}},
\label{ene12}
\end{equation}
and
\begin{equation}
 N_3=(\vert v_\chi\vert^2\vert v_\rho\vert^2+
\vert v_\chi\vert^2\vert v_\eta\vert^2 +
\vert v_\rho\vert^2\vert v_\eta\vert^2)^{\frac{1}{2}},
\label{ene3}
\end{equation}
and with the mass of the $h^0$ field given by
\begin{equation}
M^2_h=\frac A{\sqrt{2}}\left( \frac 1{\left| v_\eta \right|^2}+
\frac 1{\left| v_\chi\right| ^2}+\frac 1{\left| v_\rho \right| ^2}\right).
\label{mh}
\end{equation}

In the $CP$ even sector we have the mass matrix
\begin{equation}
\left(
\begin{array}{ccc}
-2a_2\left| v_\eta \right| ^2+\frac A{\sqrt{2}\left| v_\eta \right|
^2}\; & \; -a_6\left| v_\eta \right| \left| v_\rho \right| -\frac A{\sqrt{2}%
\left| v_\eta \right| \left| v_\rho \right| }\; &
\;-a_4\left| v_\eta \right| \left| v_\chi\right| -\frac A{\sqrt{2}\left|
v_\eta \right| \left| v_\chi\right| } \\
 & -2a_3\left| v_\rho \right| ^2+\frac A{\sqrt{2}\left| v_\rho \right|
^2} \; & \; -a_5\left| v_\rho \right| \left| v_\chi\right| -\frac A{\sqrt{2}
\left| v_\rho \right| \left| v_\chi\right| }\\
& &\; -2a_1\left| v_\chi\right| ^2+\frac A{\sqrt{2}\left| v_\chi\right| ^2}
 \end{array}\right),
\label{nsxm}
\end{equation}
in the $(X^0_\eta,X^0_\rho,X^0_\chi)^T$ basis. All those scalars are physical
but we will not write the respective mass eigenstates~\cite{mdt}. 
It is enough to stress that $X^0_a=O^H_{a i}H^0_i$ where $a=\eta,\rho,\chi$; 
$i=1,2,3$ and $O^H$ is an orthogonal $3\times 3$ matrix.


\section{Yukawa Interactions}
\label{sec:yuka}
Here we will consider the most general Yukawa interaction with real coefficients 
which is invariant under the gauge symmetry. However, there are flavor 
changing neutral currents through the neutral scalars and since these fields
get nonzero VEVs as in Eq.~(\ref{vev}) it is not straightforward to see
which phases can be absorbed by redefining the fermion fields or which 
of them, and where, survive in the lagrangian density. 

In this section we will show that it is possible to choose that all 
Yukawa interactions which have a counterpart in the standard model are $CP$ 
conserving. Hence, all $CP$ violation effects arise from the singly and/or 
doubly charged scalar-fermion interactions. The vector gauge interactions with 
the fermions are also $CP$ conserving including the singly and doubly charged 
bileptons as we will show in Sec.~\ref{sec:gq}.

\subsection{Quark-scalar interactions}
\label{subsec:qs}
First, let us consider the quark-scalar interactions. The quark multiplets
are the following
\begin{mathletters}
\label{qm}
\begin{equation}
Q_{1L}^{\prime }=\left( 
\begin{array}{c}
u_1^{\prime } \\ 
d_1^{\prime } \\ 
J_1^{\prime }
\end{array}
\right) _L\sim \left( 3,\frac 23\right) ,
\quad
Q_{mL}^{\prime }=\left( 
\begin{array}{c}
d_m^{\prime } \\ 
u_m^{\prime } \\ 
J_m^{\prime }
\end{array}
\right) _L\sim \left( 3^{*},-\frac 13\right),\;m=2,3;
\label{triq}
\end{equation}
and the respective right-handed components in singlets
\begin{equation}
U_{\alpha R}^{\prime }\sim({\bf3},{\bf1},2/3),\text{ 
}D_{\alpha R}^{\prime }\sim({\bf3},{\bf1},-1/3),
\text{ }J_{1R}^{\prime }\sim({\bf3},{\bf1},5/3); 
\text{ }J_{mR}^{\prime }\sim({\bf3},{\bf1},-4/3), 
\label{sq}
\end{equation}
\end{mathletters}
where $\alpha=1,2,3$.

With the quark multiplets in Eqs.~(\ref{qm}) and the scalar ones in 
Eq.~(\ref{e1}) we have the Yukawa terms
\begin{eqnarray}
-{\cal L}_Y&=& \overline{Q}_{1L}^{\prime }\sum_\alpha \left( G_{1\alpha
}U_{\alpha R}^{\prime }\eta+\widetilde{G}_{1\alpha }D_{\alpha
R}^{\prime }\rho \right) + 
\sum_i\overline{Q}_{iL}^{\prime }\sum_\alpha \left( F_{i\alpha }U_{\alpha
R}^{\prime }\rho ^{*}\right.
\nonumber \\ &&\mbox{}
\left.+\widetilde{F}_{i\alpha }D_{\alpha R}^{\prime
}\eta ^{*}\right) +\lambda _1\overline{Q}_{1L}^{\prime
}J_{1R}^{\prime }\chi+\sum_{i,m}\lambda _{im}\overline{Q}%
_{iL}^{\prime }J_{mR}^{\prime }\chi ^{*}+H.c.,
\label{yu1}
\end{eqnarray}
here $\alpha =1,2,3;\; i,m=2,3$
and we will assume that all coupling constants in Eq.~(\ref{yu1}) are real.
In the following subsections we will analyse, case by case, all the charged 
sectors.

\subsubsection{u-like sector--neutral scalar interactions}
\label{subsubsec:un}
After the spontaneous symmetry breaking the VEVs of the neutral scalars are 
arbitrary complex numbers, as discussed in Secs.~\ref{sec:m1}. Using the 
shifted neutral fields, given in Eq.~(\ref{vev}), from Eq.~(\ref{yu1}) we obtain
the interactions of the quarks with the neutral scalars.

In the $u$-like sector we have the interactions
\begin{equation}
-{\cal L}_Y^u=\overline{u}_{1L}^{\prime }\sum_\alpha G_{1\alpha }U_{\alpha
R}^{\prime }\eta^0+\sum_m\overline{u}_{mL}^{\prime }\sum_\alpha
F_{i\alpha }U_{\alpha R}^{\prime }\rho ^{ 0*}+H.c..
\label{yuun}
\end{equation}

In this section we will not use the freedom of choosing two phases equal to zero
because we want to see how many phases can be absorbed in the fermion fields.
Redefining the phases of the following fields~\cite{cb}
\begin{equation}
U_{\alpha R}^{\prime \prime }\equiv e^{i\theta _\eta}\, U_{\alpha R}
^{\prime },\quad
u_{mL}^{\prime \prime }\equiv e^{-i(\theta _\eta+\theta _\rho)}\,
u_{mL}^{\prime },
\label{re1}
\end{equation}
the mass matrix of the $u$-like sector become real and can be diagonalized by 
real orthogonal matrices ${\cal O}^u_{L,R}$ (recall that $\alpha=1,2,3$ and 
$m=2,3$):
\begin{equation}
\left( {\cal O}_L^u \right) ^T\Gamma ^u{\cal O}_R^u =M^u={\rm diag}\left(
m_u,m_c,m_t\right) 
\label{mmu}
\end{equation}
with
\begin{equation}
\Gamma^u=\frac 1{\sqrt{2}}\left( 
\begin{array}{ccc}
\vert v_\eta\vert G_{11} & \vert v_\eta\vert  G_{12} & \vert v_\eta\vert G_{13} \\ 
\vert v_\rho\vert F_{21} &\vert v_\rho\vert F_{22} &\vert v_\rho\vert F_{23} \\ 
\vert v_\rho\vert F_{31} &\vert v_\rho\vert F_{32} &\vert v_\rho\vert F_{33}
\end{array}
\right) 
\label{gu}
\end{equation}

Symmetry eigenstates (singly and doubly primed fields) are related to the mass 
eigenstates fields $u,c,t$ as follows:
\begin{equation}
\left( 
\begin{array}{c}
u_{1L}^{\prime } \\ 
u_{2L}^{\prime \prime } \\ 
u_{3L}^{\prime \prime }
\end{array}
\right) ={\cal O}_L^u \left( 
\begin{array}{c}
u \\ 
c \\ 
t
\end{array}
\right) _L,\qquad \left( 
\begin{array}{c}
U_{1R}^{\prime \prime } \\ 
U_{2R}^{\prime \prime } \\ 
U_{3R}^{\prime \prime }
\end{array}
\right) ={\cal O}_R^u \left( 
\begin{array}{c}
u \\ 
c \\ 
t
\end{array}
\right) _R 
\label{mou}
\end{equation}

The Yukawa interaction in Eq.~(\ref{yuun}) can be written as
\begin{eqnarray}
-{\cal L}_Y^u&=&\overline{U}_L M^u U_R+ 
\,\overline{U}_L M^u U_R\, \frac{X^0_\rho -
iI^0_\rho}{\vert v_\rho\vert}
\nonumber \\ &&\mbox{}+\overline{U}_L({\cal O}^u_L)^T \Delta \text{ }{\cal 
O}_L^u M^u{U}_R\left( \frac{ X^0_\eta+
iI^0_\eta}{\vert v_\eta\vert} -\frac{X^0_\rho-iI^0_\rho}{\vert v_\rho\vert }
\right) +H.c., 
\label{yuunf}
\end{eqnarray}
with 
\begin{equation}
\Delta \equiv \left(
\begin{array}{ccc}
1 & 0 & 0 \\ 
0 & 0 & 0 \\ 
0 & 0 & 0
\end{array}
\right).
\label{delta}
\end{equation}

Since there is no mixing among $X$'s and $I$'s we have no $CP$ violation in 
this sector.

\subsubsection{$d$-like sector--neutral scalar interactions}
\label{subsubsec:dn}
Similarly, in the $d$-like sector, the interaction with the neutral Higgs 
scalars are
\begin{equation}
-{\cal L}_Y^d=\overline{d}_{1L}^{\prime }\sum_\alpha \widetilde{G}_{1\alpha
}D_{\alpha R}^{\prime }\rho ^0
+\sum_m\overline{d}_{mL}^{\prime }\sum_\alpha 
\widetilde{F}_{m\alpha }D_{\alpha R}^{\prime }\eta ^{0*}+H.c..
\label{yudn}
\end{equation}

Making the following phase redefinition
\begin{equation}
D_{\alpha R}^{\prime \prime }\equiv e^{i\theta _\rho }\,
D_{\alpha R}^{\prime } ,\quad d_{mL}^{\prime \prime }\equiv 
e^{-i(\theta _\rho +\theta _\eta)}\,
d_{mL}^{\prime },
\label{re2}
\end{equation}
we obtain a real mass matrix which can be diagonalized with a orthogonal 
transformation
\begin{equation}
\left( {\cal O}_L^d\right) ^T\Gamma ^d{\cal O}_R^d=M^d={\rm diag}\left(
m_d,m_s,m_b\right) 
\label{e2}
\end{equation}
with
\begin{equation}
\Gamma ^d\equiv \frac 1{\sqrt{2}}\left(
\begin{array}{ccc}
\vert v_\rho\vert \widetilde{G}_{11} &\vert v_\rho\vert \widetilde{G}_{12} & 
\vert v_\rho\vert \widetilde{G}%
_{13} \\ 
\vert v_\eta\vert \widetilde{F}_{21} & \vert v_\eta\vert\widetilde{F}_{22} & 
\vert v_\eta\vert\widetilde{F}_{23} \\ 
\vert v_\eta\vert\widetilde{F}_{31} & \vert v_\eta\vert\widetilde{F}_{32} & 
\vert v_\eta\vert \widetilde{F}_{33}
\end{array}
\right)
\label{e3}
\end{equation}

The symmetry eigenstates (singly and doubly primed fields) are related to the 
mass eigenstates (unprimed fields) as follows:
\begin{equation}
\left( 
\begin{array}{c}
d_{1L}^{\prime } \\ 
d_{2L}^{\prime \prime } \\ 
d_{3L}^{\prime \prime }
\end{array}
\right) ={\cal O}_L^d\left( 
\begin{array}{c}
d \\ 
s \\ 
b
\end{array}
\right) _L,\qquad \left( 
\begin{array}{c}
D_{1R}^{\prime \prime } \\ 
D_{2R}^{\prime \prime } \\ 
D_{3R}^{\prime \prime }
\end{array}
\right) ={\cal O}_R^d\left( 
\begin{array}{c}
d \\ 
s \\ 
b
\end{array}
\right)_R,  
\label{e5}
\end{equation}
and the interactions in Eq.~(\ref{yudn}) become
\begin{eqnarray}
-{\cal L}_Y^d&=&\overline{ D}_LM^d{D}_R+
\overline{D}_LM^d{D}_R\,
\frac{X^0_\eta-iI^0_\eta}{\vert v_\eta\vert} 
\nonumber \\ &&\mbox{}
+\overline{D}_L\left( {\cal O}_L^d\right) ^T\triangle \text{ }
{\cal O}_L^dM^d{D}_R\left( \frac{X^0_\rho
+iI^0_\rho}{\vert v_\rho\vert}-
\frac{X^0_\eta-iI^0_\eta}{\vert v_\eta\vert}
\right) +H.c. .
\label{e6}
\end{eqnarray}

Again, we see that if $X$'s and $I$'s do not mix among them we have not $CP$ 
violation through the exchange of neutral fields. This is the case of the
model with only three triplets in Sec.~\ref{sec:m1}. On the other
hand, we have this type of $CP$ violation in the model with three triplets and 
one sextet~\cite{laplata1}.

\subsubsection{Exotic quarks--neutral scalar interactions}
\label{subsubsec:en}
In the sector involving exotic quarks we have
\begin{equation}
-{\cal L}_Y^J=\lambda _1\overline{J}_{1L}^{\prime }J_{1R}^{\prime }\chi^0
+\sum_{i,m}\lambda _{im}\overline{J}_{iL}^{\prime }J_{mR}^{\prime }\chi^{0*}
 +H.c. 
\label{e7}
\end{equation}

Making the redefinition of the right-handed components of the exotic quarks
\begin{equation}
J_{1R}^{\prime \prime }\equiv e^{ i\theta _\chi}\, J_{1R}^{\prime },
\quad
J_{mR}^{\prime \prime }\equiv e^{ -i\theta _\chi}\, J_{mR}^{\prime}, 
\label{re3}
\end{equation}
we have
\begin{equation}
-{\cal L}_Y^J=\lambda _1\overline{J}_{1L}^{\prime }J_{1R}^{\prime \prime }%
\frac{\vert v_\chi\vert}{\sqrt{2}}\left( 1+\frac 
{X^0_\chi+iI^0_\chi}{\vert v_\chi\vert}\right) + 
\sum_{i,m}\lambda _{im}\overline{J}_{iL}^{\prime }J_{mR}^{\prime \prime }%
\frac{\vert v_\chi\vert}{\sqrt{2}}\left( 1+\frac{X^0_\chi-
iI^0_\chi}{\vert v_\chi\vert} \right) +H.c. .
\label{e8}
\end{equation}

Notice that $J^\prime_1$ (or $J^{\prime\prime}_1$) does not mix with any other 
quarks but $J^\prime_{2,3}$ (or $J^{\prime\prime}_{2,3}$) mix 
between themselves since they have the same charge. Here we use, when
necessary, 
\begin{equation}
J\equiv J^{\prime\prime}_1,\quad \mbox{and}\quad 
\left(J^{\prime\prime}_{L,R}\right)_m=({\cal O}^J_{L,R})_{mi}(j_{L,R})_i,
\label{js}
\end{equation}
where $J$ and $j_n$ with $n=1,2$ denote mass eigenstates, {\it i.e.}, the mass 
eigenstates in the exotic quark sector will be denoted when necessary $J$ for 
the charge 5/3 quark and $j_{1,2}$ for the two charge $-4/3$ quarks.  
Hence Eq.~(\ref{e8}) can be written in terms of mass eigenstates exotic 
fermions
\begin{equation}
-{\cal L}_Y^J=m_J\bar{J}_LJ_R\left( 1+\frac 
{X^0_\chi+iI^0_\chi}{\vert v_\chi\vert}\right) + 
\bar{j}_LM^Jj_R \left( 1+\frac{X^0_\chi-
iI^0_\chi}{\vert v_\chi\vert} \right) +H.c.,
\label{e8b}
\end{equation}
where $M^J=\mbox{diag}(m_{j_i},m_{j_2})$.

\subsubsection{Singly charged scalar--quark interactions}
\label{subsubsec:chq}
The interaction lagrangian is
\begin{equation}
-{\cal L}_Y^{u-d}=\sum_\alpha \left( \overline{d}_{1L}^{\prime }G_{1\alpha
}U_{\alpha R}^{\prime }\eta _1^{-}+\overline{u}_{1L}^{\prime }
\widetilde{G}_{1\alpha }D_{\alpha R}^{\prime }\rho ^{ +}\right) 
+ \sum_{i,\alpha} \left( \overline{d}_{iL}^{\prime }F_{i\alpha }U_{\alpha
R}^{\prime }\rho ^{ -}+\overline{u}_{iL}^{\prime }\widetilde{F}%
_{i\alpha }D_{\alpha R}^{\prime }\eta _1^{ +}\right)+H.c.
\label{yuca1}
\end{equation}

Using the phase redefinition of Eqs.~(\ref{re1}), (\ref{re2}) and (\ref{re3})
in Eq.~(\ref{yuca1}) we have
\begin{eqnarray}
-{\cal L}_Y^{u-d}&=&\frac{\sqrt2}{\vert v_\rho\vert}\,
\overline{D}_L\left( {\cal O}_L^d\right) ^TK_1{\cal O}_L^uM^u U_R\, 
\rho ^{-}+\overline{U}_L\left( {\cal O}_L^u\right) ^T
K_2{\cal O}_L^dM^d{ D}_R
\frac{\sqrt{2}}{\vert v_\eta\vert}\, \eta _1^{+}
\nonumber \\ &&\mbox{}
+\overline{D}_L\left( {\cal O}_L^d\right) ^TK_1\triangle \text{ }{\cal O}_L^u
M^u U_R  \left[\frac{\sqrt{2}}{\vert v_\eta\vert}\, \eta_1^{-}-
\frac{\sqrt{2}}{\vert v_\rho\vert }\, \rho ^{ -}\right]
\nonumber \\ &&\mbox{}+ 
\overline{U}_L\left( {\cal O}_L^u\right) ^TK_2\triangle \text{ }{\cal O}_L^dM^d 
D_R\left[ \frac{\sqrt{2}}{\vert v_\rho\vert }\, 
\rho ^{+}- \frac{\sqrt{2}}{\vert v_\eta\vert}\,\eta _1^{ +}\right] 
+H.c. , 
\label{yucud}
\end{eqnarray}
being
\begin{equation}
K_1={\rm diag}( e^{-i\theta _\eta},e^{ i\theta _\rho},e^{i\theta_\rho}), \;\;
\quad {\rm and} \quad
K_2={\rm diag}(e^{-i\theta _\rho},e^{i\theta _\eta}, e^{i\theta _\eta}).
\label{k12}
\end{equation}

Notice that if we consider the more general potential, {\it i.e.}, with
the $a_{10}$ term, we have, according to the Eq.~(\ref{scm}) a general
mixing among all singly charged scalars. There is $CP$ violation through
the exchange of a singly charged scalar in this case. However, when we
consider the case when $a_{10}=0$ the mixing in that sector is 
given by Eqs.~(\ref{scf1}) and (\ref{scf2}) and there is no $CP$ violation
in the interaction of Eq.~(\ref{yucud}) if we also chose that $\theta_\eta=
\theta_\rho=0$ by making a $SU(3)$ transformation.

Finally, we have the interaction involving the exotic quarks,
\begin{eqnarray}
-{\cal L}_Y^{J-\chi }&=&\sum_\alpha \bar{J}_L G_{1\alpha}
{\cal O}^{u}_{R\alpha\beta}
U_{\beta R}\eta^+_2e^{-i\theta_\eta}+
\frac{m_J\sqrt2}{\vert v_\chi\vert}\,\bar{U}_{L}\left({\cal O}^u_L\right)^T
\Delta J_R\,e^{-i\theta _\chi} \chi ^{-}\nonumber \\ &+&
\frac{\sqrt2}{\vert v_\chi\vert}\,
\bar{D}_{L}\left({\cal O}^{d}_L\right)^T \overline{\Delta}
\left({\cal O}^J_R\right)^TM^Jj_{R}\,
e^{-i(\theta_\rho +\theta _\eta-\theta _\chi)} \chi^{+}+H.c., 
\label{yue}
\end{eqnarray}
where we have already used Eqs.~(\ref{re1}) and (\ref{re2}) and defined 
\begin{equation}
\overline{\Delta} \equiv \left(
\begin{array}{ccc}
0& 0 & 0 \\ 
0 & 1 & 0 \\ 
0 & 0 & 1
\end{array}
\right).
\label{deltab}
\end{equation}
The phases of the fields $u_{1L}^{\prime }$ and $d_{1L}^{\prime}$
(and those of the $J_{1L}^{\prime }$ and  $J_{mL}^{\prime }$ too) are, 
in principle, still free. 
However, we can not change the phases of these fields 
unless we change the phases in the Eqs.(\ref{yuun}) and (\ref{yudn}) (or also 
in Eqs.(\ref{yucud})). Notwithstanding,
the phases in those equations have already being absorbed after the 
redefinition given in Eqs.~(\ref{re1}) and (\ref{re2}), respectively.
We can not absorb phases any more so we have $CP$ violation through the 
exchange of the singly charged scalars which are coupled to the exotic and
known quarks. 
Hence, we have shown that the only
phase redefinition in the quark fields are those in Eqs.~(\ref{re1}), 
(\ref{re2}) and (\ref{re3}). It means that the observable $CP$ violating phase
is ${\rm Im} v_\chi$ (or $-{\rm Im}\alpha$). 
This will be also the case in the interactions with the doubly charged scalars.

\subsubsection{Doubly charged scalar--quark interactions}
\label{subsubsec:dch}
The doubly charged Higgs scalars couple the known quarks with the exotic 
ones and by using Eqs.~(\ref{re1}) and (\ref{re2}) we have
\begin{mathletters}
\label{dc}
\begin{equation}
-{\cal L}_Y^{J-\rho}=\sum_\alpha \bar{J}_{L}G_{1\alpha} 
{\cal O}^D_{R\alpha\beta}
D_{\beta R}\,e^{-i\theta_\rho}\rho^{++}+
\sum_{i,l,\alpha,\beta}\overline{j}_{Ll}\left({\cal O}^J_{L}\right)^T_{li} 
F_{i\alpha}  {\cal O}^u_{R\alpha\beta}U_{\beta R}\,
e^{-i\theta_\eta} \rho^{--}+H.c.,
\label{dch1}
\end{equation}
and
\begin{equation}
-{\cal L}_Y^{J-\chi}=\frac{m_J\sqrt2}{\vert v_\chi\vert}\,
\bar{D}_{L}\left({\cal O}^d_L\right)^T\Delta
J_{R}\,e^{-i\theta_\chi}\chi ^{--}+
\frac{\sqrt2}{\vert v_\chi\vert}\,\bar{U}_{L}\overline{\Delta}
\left({\cal O}^u_L\right)^T{\cal O}^J_LM^Jj_{R}\,
e^{i(\theta _\chi-\theta _\rho -\theta_\eta)} \chi ^{++}+H.c.. 
\label{dc2}
\end{equation}
\end{mathletters}

From the same argument of the preceeding subsections we can see that there is 
$CP$ violation in the doubly charged Higgs exchange too. The fields $\rho^{++}$ 
and $\chi^{++}$ are given in terms of the respective mass eigenstates in 
Eq.~(\ref{dcf}). Fermion fields are all the mass eigenstates.

\subsection{Lepton--scalar interactions}
\label{subsec:leptons}
Now, let us consider the leptonic sector. 
The leptons are assigned to the following representations: 
\begin{equation}
\Psi_{aL}=\left(\begin{array}{c}
\nu_{l_a} \\  l'^-_a\\ E'^+_a
\end{array}\right)_L \sim ({\bf3},0);\quad l'^-_{aR}\sim({\bf1},-1),\;
E'^-_{aR}\sim({\bf1},+1),\;\;a = e, \, \mu , \, \tau.
\label{lep1}
\end{equation}
The Yukawa interactions in this sector are
\begin{equation}
-{\cal L}^l_Y=G^e_{ab}\overline{\psi }_{aL}l_{bR}^{\prime -}\,\rho
+G^\psi_{ab}\overline{\psi }_{aL}E_{bR}^{\prime +}\,\chi+H.c. .
\label{yunm1}
\end{equation}
Next, we will consider each type of interactions as we did in the quark sectors.

\subsubsection{Lepton--neutral scalar interactions}
\label{subsubsec:yul1}
In this sector the relation between
the symmetry eigenstate fields ($E'_{a},\,l'_{aL};\;a=e,\mu,\tau$) 
and the mass eigenstate fields ($l_i=e,\mu,\tau;\,E_i=E_1, E_2, E_3$) 
is obtained through orthogonal matrices (note that in the leptonic sector 
$i=1,2,3$)
\begin{equation}
l'_{aL}={\cal O}^e_{Lai}l_{iL},\quad
l'_{aR}={\cal O}^e_{Rai}l_{iR},\quad
E'_{aL}={\cal O}^E_{Lai}E_{iL},\quad
E'_{aR}={\cal O}^E_{Rai}E_{iR},
\label{redef}
\end{equation}

The respective mass matrices are defined as follows:
\begin{equation}
M^e=\frac{\vert v_\rho\vert}{\sqrt2}\,\left({\cal O}^{e}_L\right)^T\,G^e\,
\,{\cal O}^{e}_R,\quad M^E=\frac{\vert v_\chi\vert}{\sqrt2}\,
\left({\cal O}^E_L\right)^T \,G^E\,{\cal O}^E_R,
\label{lmass}
\end{equation}
with $M^e={\rm diag}(m_e,m_\mu,m_\tau)$ and $M^E=
{\rm diag}(m_{E_1},m_{E_2},m_{E_3})$. Notice that the exotic lepton masses are 
proportional to the larger VEV, $v_\chi$.
We see that there is no $CP$ violation in this sector too.

In terms of the physical lepton fields we have
\begin{equation}
-{\cal L}^l_Y=\bar{l}_{eL}M^el_{eR}+
\bar E _{L}M^E E_{R}+
\bar l_{eL}M^el_{eR}\,\frac{ X^0_\rho +iI^0_\rho}{\vert v_\rho\vert}
+\bar E_L M^E E_{R}\,\frac{X^0_\chi+
iI^0_\chi}{\vert v_\chi\vert}
+H.c., 
\label{lyum1}
\end{equation}
and we have redefined the right-handed lepton components
\begin{equation}
l^{\prime\prime}_{iR}= e^{i\theta_\rho}l_{iR},\;\;
E^{\prime\prime}_{iR}=e^{i\theta_\chi}E_{iR},
\label{flrh}
\end{equation}
but however we have omitted the double primed in Eq.~(\ref{lyum1}). 

\subsubsection{Lepton--singly charged scalar interactions}
\label{subsubsec:yul2}
The interactions involving singly charged scalars are
\begin{equation}
-{\cal L}_Y^{\nu -l,\psi}=\frac{\sqrt{2}}{\vert v_\rho\vert}\;
\overline{\nu'_L}M^el_R
\rho ^{ +}+ 
\frac{\sqrt{2}}{\vert v_\chi\vert }\;\overline{\nu'_L}
{\cal K}M^E E_{R}e^{ -i(\theta _\chi
-\theta _\rho)} \chi^{ -}+H.c.,
\label{hclm1}
\end{equation}
where we have defined
\begin{equation}
\overline{\nu'_L}=\overline{\nu_L}{\cal O}^e_L e^{-i\theta_\rho},
\label{nure}
\end{equation}
and ${\cal K}\equiv \left( {\cal O}_L^e\right) ^T{\cal O}^E_L$.
We see that even if we choose $\theta_\rho=0$ the phase $\theta_\chi$ survives 
and we have $CP$ violation by the singly charged scalar exchanging.  

\subsubsection{Lepton--doubly charged scalar interactions}
\label{subsubsec:yul3}
The interactions with the doubly charged scalars in the lepton sector are given 
by
\begin{equation}
-{\cal L}_Y^{l-E}=\frac{\sqrt{2}}{\vert v_\rho\vert}\,\bar E_{L}
{\cal K}^T M^el_{R}
e^{-i\theta _\rho} \rho ^{++}+ 
\frac{\sqrt{2}}{\vert v_\chi\vert }\,\overline{l_{L}} {\cal K}M^E
E_{R}e^{-i\theta _\chi}\chi ^{--}+H.c. .
\label{dcyul1}
\end{equation}
As in the previous case we have $CP$ violation in the doubly charged scalar 
sector even if we choose $\theta_\rho=0$. 

\section{Gauge interactions}
\label{sec:gq}
Next, we will verify in what conditions all phase redefinitions that have been 
done in the previous section do not appear in the vector boson-fermion 
interactions.

\subsection{Quark--vector boson interactions}
\label{subsec:qgi}
With the redefinition of the phases in Eqs.~(\ref{re1}) and (\ref{re2})
the mixing matrix in the charged currents coupled to the $W^\pm$ is real 
\begin{equation}
{\cal L}_Y^{W-q}=-\frac g{\sqrt{2}}\overline{U}_L\gamma^u
V_{\rm CKM}D_LW_\mu ^{+}+H.c., 
\label{wq}
\end{equation}
with the CKM matrix defined as $V_{\rm CKM}=
\left( {\cal O}_L^u\right) ^T{\cal O}_L^d$.
So we have no $CP$ violations in this sector. Similarly, we have in the
charged currents coupled to the vector bileptons $V^+$ and $U^{--}$,
\begin{equation}
{\cal L}_Y^{V-q}=-\frac g{\sqrt{2}}\left( \overline{J}_{1L}^{\prime }\gamma
^\mu u_{1L}^{\prime }-\sum_{i,m}\overline{d}_{iL}^{\prime \prime }\gamma
^\mu J_{mL}^{\prime }\right) e^{ -i(\theta _\rho +\theta _\eta)}
V_\mu ^{+}+H.c. ,
\label{vq}
\end{equation}
and
\begin{equation}
{\cal L}_Y^{U-q}=-\frac g{\sqrt{2}}\left( \overline{J}_{1L}^{\prime }\gamma
^\mu d_{1L}^{\prime }-\sum_{i,m}\overline{u}_{iL}^{\prime \prime }\gamma
^\mu J_{mL}^{\prime }\right) e^{-i(\theta _\rho +\theta _\eta)}
U_\mu ^{--}+H.c. .
\label{uq}
\end{equation}
However, we always can choose $v_\eta=v_\rho=0$ by using a $SU(3)$ 
transformation. Hence, we will not have $CP$ violation in the bilepton 
sector.

\subsection{Lepton--vector boson interactions}
\label{subsec:gl}
The charged current interactions with the vector bosons in the leptonic 
sector are
\begin{equation}
{\cal L}^{CC}_l=-\frac{g}{\sqrt2}\,\sum_a\left( \bar{\nu}_{aL}\gamma^\mu
l_{aL}W^+_\mu+\bar{\psi}_{aL}\gamma^\mu \nu_{aL}V^+_\mu
+\bar{\psi}_{aL}\gamma^\mu l_{aL}U^{++}_\mu+H.c.\right),
\label{gl1}
\end{equation}
where all fields are still symmetry eigenstates (but we have omitted the prime).
Then, the charged current interactions in terms of the 
physical basis, using Eq.~(\ref{redef}), is given by
\begin{equation}
{\cal L}^{CC}_l=-\frac{g}{\sqrt2}\left(\bar\nu_L\gamma^\mu l_LW^+_\mu+ 
\bar \psi_L\gamma^\mu {\cal K}^T  \nu_L V^+_\mu\,e^{i\theta_\rho}-
\bar \psi_L \gamma^\mu{\cal K}l_L U^{++}_\mu+H.c.\right),
\label{cc331}
\end{equation}
where ${\cal K}\equiv \left( {\cal O}_L^e\right) ^T{\cal O}_L^\psi$ and we have 
used the redefinition of the neutrino fields in Eq.~(\ref{nure}).
Although a phase appears we can always choose it as being zero by an
$SU(3)$ transformation.

We see from Eq.~(\ref{cc331}) that for massless neutrinos we have no mixing
in the charged current coupled to $W^+_\mu$ but we still have mixing 
in the charged currents coupled to $V^+_\mu$ and $U^{++}_\mu$. 
That is, if neutrinos are massless we can always choose 
$\bar\nu'_L\equiv \bar\nu_L{\cal O}^e_L$. However,  
the charged currents coupled to $V^+_\mu$ and $U^{++}_\mu$ are not diagonal in 
flavor space since the mixing matrix ${\cal K}$ survives.
Thus, there is not $CP$ violation in this sector if we use the freedom to 
choose $\theta_\rho=0$.

The pure gauge boson interactions also conserve $CP$ if the $SU(3)_L$ gauge 
and $U(1)_N$ vector bosons $W^a\;\;\mbox{with}\,a=1,...8$,
and $B$, respectively transform as 
\begin{equation}
(W^1_\mu,W^2_\mu W^3_\mu,W^4_\mu,W^5_\mu,W^6_\mu,W^7_\mu,W^8_\mu,B_\mu)
\stackrel{CP}{\to} -(W^{1\mu},-W^{2\mu},W^{3\mu},-W^{4\mu},W^{5\mu},-W^{6\mu},
W^{7\mu},W^{8\mu},B^\mu).
\label{e12}
\end{equation}
It means that the physical fields transform as 
\begin{equation}
(W^+_\mu, V^+_\mu,U^{++}_\mu,A_\mu,Z_\mu,Z^\prime_\mu)
\stackrel{CP}{\to}-(W^{\mu-}, -V^{\mu-},-U^{\mu--},A^\mu,Z^\mu,Z^{\prime\mu}),
\label{wvu}
\end{equation}

We have shown that if all term, except the trilinear term in the scalar 
potential conserve $CP$ this symmetry will be broken when the neutral scalars
gain a complex VEV. 

\section{Phenomenological consequences}
\label{sec:feno}
As it is well known, the violation of the $CP$ symmetry was discovered in 
1964 in the $K^0-\bar{K}^0$ system~\cite{cpexp}. Up to now, it is only in this
particular system in which $CP$ violating effects has been seen. If the source 
of the $CP$ violation is the weak interactions we expect also to see its
effects in $B$ decays. However, only a general discussion is presented here
concerning the mesons case~\cite{cp6}. 
On the other hand a detailed study of the electric dipole 
moments of the neutron and charged leptons is shown. 

\subsection{Quark sector}
\label{sec:fenoq}
\subsubsection{$CP$ violation in the neutral meson systems}
\label{subsubsec:kkbar}
In Fig.~1 we show the tree level contributions to the mass difference 
$\Delta M_K=2{\rm Re}M_{12}$ (where $M_{12}=\langle K^0\vert{\cal H}_{eff}\vert 
\bar K^0\rangle$). 
These diagrams exist because of the flavor changing neutral 
currents in Eq.~(\ref{e6}). The $H^0$'s contributions to $\Delta M_K$ have been 
consider in Ref.~\cite{laplata2}. For $m_H\sim150$ GeV the constraint coming 
from the experimental value of $\Delta M_K$ implies 
$({\cal O}^d_L)_{11}({\cal O}^d_L)_{12}\lesssim0.01$.
There are also tree level contributions to 
$\Delta M_K$ coming from the $Z'$ exchange which were considered in 
Ref.~\cite{dumm}. Similarly with the mass difference of $B^0_d-\bar{B}^0_d$,
$B^0_s-\bar{B}^0_s$ and $D^0-\bar{D}^0$ systems.
However, in the present model, the $CP$ violating parameters
like $\varepsilon_K$ have only contributions coming
from box diagrams involving the one or two doubly charged and one singly 
charged scalars as can be seen from Fig.~2. The direct $CP$ violation parameter 
$\varepsilon^\prime_K$ has contributions at the 1-loop level too, as is shown 
in Fig.~3. However, there is no penguin contributions as we will show later. 

Although we will not make here a detailed calculation of $\varepsilon_K$ and 
$\varepsilon'_K$ and the respective parameters in the $B^0_d-\bar{B}^0_d$
and $B^0_s-\bar{B}^0_s$ systems we can notice that in principle the model can 
give values for these $CP$ violating parameters which are in accord with data 
since they depend on different mixing matrices 
${\cal O}^{u,d}_{L,R}$. From Eqs.~ (\ref{dch1}) and (\ref{dc2}) we see that 
$\varepsilon_K$ comes from diagrams like those shown in Fig.~2(a) and 2(b). 
(There are two other diagrams as the ones in Figs. 2 but with the lines of $Y$ and $J,j_{1,2}$
interchanged.) There are diagrams similar to those in Figs.~2 but with one of 
the scalar bileptons being changed by a vector bilepton and with no
mass insertion. This contributions are less suppressed by the mixing angles
but for the vector bilepton mass. 

Similar diagrams do exist for the other $B$ mesons. The
coefficient of the amplitude produced by diagrams like that in Fig.~2
are proportional to the several mixing matrix elements and Yukawa couplings.
For instance, for neutral $K$ system, from Fig.~2(a), up to a $\sin(4\theta_\alpha)$ factor the 
amplitude is proportional to
\begin{mathletters}
\label{fenopa}
\begin{equation}
G_{1\alpha}\left({\cal O}^d_R\right)_{\alpha1}
\left({\cal O}^d_L\right)_{21}
G_{1\beta}\left({\cal O}^d_R\right)_{\beta1}
\left({\cal O}^d_L\right)_{11},
\label{kk}
\end{equation}
while for $B$ mesons, similar diagrams to that in Fig.~2(a) imply that the
amplitudes are proportional to
\begin{equation}
G_{1\alpha}\left({\cal O}^d_R\right)_{\alpha1}
\left({\cal O}^d_L\right)_{31}
G_{1\beta}\left({\cal O}^d_R\right)_{\beta1}
\left({\cal O}^d_L\right)_{11},
\label{bb}
\end{equation}
for the neutral $B_d$ system and,
\begin{equation}
G_{1\alpha}\left({\cal O}^d_R\right)_{\alpha2}
\left({\cal O}^d_L\right)_{31}
G_{1\beta}\left({\cal O}^d_R\right)_{\beta3}
\left({\cal O}^d_L\right)_{21},
\label{bsbs}
\end{equation}
\end{mathletters}
for the neutral $B_s$ system. We see from Eqs.~(\ref{fenopa}) that the 
orthogonality condition implies that if we have chosen two of the 
$\epsilon_{K,B_d,B_s}$ parameters the third one is fixed.  
A similar analyse follows from Fig.~2(b) and the equivalent diagrams for the
$B$ systems. 

In the $D$ mesons case the mixing matrix is different since in these models the 
left-handed mixing matrix ${\cal O}^u_L$ survives in different places of the 
lagrangian so it is not natural to set them equal to zero (see below).
Hence, we see that all $CP$ phenomenology in the meson systems can be 
accommodated, in principle, in the present model.

Concerning the direct $CP$ violating parameters $\epsilon'$ its contribution
come from diagrams like the one in Fig.~3. The vertices are given in 
Eqs.~(\ref{yue}) and (\ref{dc2}).
There that there are a GIM-like cancellation between the 
contributions of $j_1$ and $j_2$. It means that the suppression of $\epsilon'$
does not give a strong constraint on the masses of $j_{1,2}$.
There are similar diagrams with the $Y^-$
substituted by a $V^-$ vector bilepton and mass insertions in the exotic
quarks lines. 
As in the case of the $\epsilon$ parameters, once we have chosen the 
appropriate value of the mixing matrix elements (and Yukawa couplings) for 
explaining the observed value in the $K^0-\bar{K}^0$ system, the $\epsilon$ 
related to the other $B$ or $B_s$ systems is at least of the same order of 
magnitude and the third one rather small. This is as expected since this sort 
of model has a superweak character. More details will be given 
elsewhere~\cite{cp6}.

\subsubsection{Electric dipole moment}
\label{subsubsec:edmn}
It is well known that the discovery of a non-zero electric dipole moment (EDM) 
for the neutron (or another elementary non-degenerate system) would be a direct 
evidence for both $CP$ and $T$ violation. The current experimental upper bound 
is~\cite{pdg}
\begin{equation}
\vert d_n\vert<1.1\times 10^{-25}\,e\;{\rm cm}.
\label{edmn}
\end{equation}
In the standard model the neutron electric dipole moment (EDM) arises at the 
three-loop level and for this reason is very small, $\sim10^{-34}$ $e$ 
cm~\cite{czar}.

In the present model we can calculate the EDM of the $d$ and $u$ quarks at the 
1-loop level. In principle the contributions are those shown in Figs.~4 and 5. 
However, we can see from the interactions in Eqs.~(\ref{yucud}), (\ref{yue}), 
(\ref{dch1}), and (\ref{dc2}) with the phase convention 
$\theta_\eta=\theta_\rho=0$ and the coefficient $a_{10}=0$, that neither the 
diagram in Fig.~4(a) nor that in 4(b) contribute to the EDM of the quark $d$, 
thus we have $d_d=0$, at this level of approximation. 
For the quark $u$ we have not only the diagram in Fig.~5(a) but also the one in 
Fig.~5(b). We obtain (recall that $\theta_\chi=-\theta_\alpha$)
\begin{equation}
d_u=\frac{em^2_um_J}{
32\pi^2(\vert v_\chi\vert^2+\vert v_\eta\vert^2)m^2_{Y^+}}
\;\left({\cal O}^u_{L11}\right)^2\;F(m_u,m_J)\,
\sin(2\theta_\alpha),
\label{edmq2}
\end{equation}
where 

\begin{equation}
F(m_u,m_J)\!\!=\!\!-\frac{m^2_{Y^+}}{2m_u^2}\ln\frac{m^2_{Y^+}}{m^2_J}
+\frac{m^2_{Y^+}} {2m_u^2\Delta_u}(m^2_{Y^+}+m^2_u-m^2_{J}) \,
\ln\left[ \frac{m^2_J+m^2_{Y^+}-m^2_u+\Delta_u}
{m^2_J+m^2_{Y^+}-m^2_u-\Delta_u}\right],
\label{fsqu}
\end{equation}
and
\begin{equation}
\Delta^2_u=(m^2_{Y^+}+m^2_J-m^2_u)(m^2_{Y^+}-m^2_J-m^2_u).
\label{deltaq1}
\end{equation}

The measured parameter is the EDM of the neutron which in the quark model
is given in terms of the constituent quarks's EDM:
\begin{eqnarray}
d_n&=&\frac{4}{3}d_d-\frac{1}{3}d_u=-\frac{1}{3}\,d_u\nonumber \\ &&\mbox{}
\simeq -1.3\times 10^{-22}\,\left({\cal O}^u_{L11}\right)^2\;
\sin(2\theta_\alpha)\quad e\;{\rm cm},
\label{edmq3}
\end{eqnarray}
where we have made the approximation $\vert v_\eta\vert\approx 
\vert v_\rho\vert$ in order to use the doubly charged vector bilepton mass 
$M_U$, given by $M^2_U=(g^2/4)(\vert v_\chi\vert^2+\vert v_\rho\vert^2)$, 
instead of $\vert v_\chi\vert^2+\vert v_\eta\vert^2$ in Eq.~(\ref{edmq2});
and $G_F/\sqrt2=g^2/8M^2_W$. We have used $M_U=300$ GeV,
$M_{Y^{+}}=100$ GeV; $m_J=50$ GeV and $m_u=0.002$ GeV. However,
the value of $d_u$ (or $d_n$) is not sensible to the masses of the exotic
particles $m_{Y^{+}}$, $M_U$ and $m_J$, at least with $M_U$ lesser than 10 TeV.
It means that only the mixing element $\left({\cal O}^u_{L11}\right)^2$, for any
value of $\sin(2\theta_\alpha)$, have to be invoked in order to obtain an
EDM of the neutron compatible with the data in Eq.~(\ref{edmn}). This is not
in conflict with the CKM mixing matrix in the charged
current coupled to the vector boson $W^\pm$ since the later is defined as
$V_{\rm CKM}=\left( {\cal O}^u_L\right) ^T{\cal O}^d_L$. Since
in Eq.~(\ref{edmq3}) only the matrix element ${\cal O}^u_{L11}$ related to 
the $u$-like quarks appears and,
with the phase convention used here, the mixing matrices related
to the $d$-like quarks do not enter at all in this sort of models and we cannot
use $V_{\rm CKM}={\cal O}^d_L$ as is usually done in the standard electroweak 
model.
 
In fact, the matrices ${\cal O}^{u,d}_L$ will appear in the neutral currents 
coupled to the extra neutral vector boson $Z^{\prime0}$. We have
\begin{eqnarray}
{\cal L}_{Z^\prime}&=&-\frac{g}{2c_W}
\left[\bar{U}_L\gamma^\mu\, {\cal O}^u_L\,Y^U_L(Z^\prime)\,{\cal O}^u_L\,U_L+ 
\bar{U}_R\gamma^\mu 
\,
{\cal O}^u_R\,Y^U_R(Z^\prime) {\cal O}^u_R\,U_R\right.\nonumber \\ &+&
\left.\bar{D}_L\gamma^\mu\, {\cal O}^d_L\,Y^D_L(Z^\prime)\,{\cal O}^d_L\,D_L+ 
\bar{D}_R\gamma^\mu \,
{\cal O}^d_R\,Y^D_R(Z^\prime)\,{\cal O}^d_R\,D_R\right] Z^\prime_\mu,
\label{nczp}
\end{eqnarray}
with the matrices
\begin{equation}
Y^U_L(Z')=Y^D_L(Z')=-\frac{1}{\sqrt{3}h(s_W)}\left(
\begin{array}{ccc}
1\,\, & \,\,       0\,\,   &\,\,    0 \\
0\,\, &\,\, 2s_W-1\,\, & \,\,   0  \\
0\,\, & \,\,       0\,\,   &\,\,  2s_W-1
\end{array}
\right)
\label{yl}
\end{equation}
and
\begin{equation}
Y^U_R(Z')=-\frac{4s_W}{\sqrt{3}h(s_W)}\left(
\begin{array}{ccc}
1\,\, &\,\,        0 \,\,  & \,\,   0 \\
0\,\, &\,\,        1\,\,  & \,\,   0  \\
0\,\, &\,\,        0\,\,   &\,\,   1
\end{array}
\right), \qquad Y^D_R(Z')=\frac{2s_W}{\sqrt{3}h(s_W)}\left(
\begin{array}{ccc}
1\,\, &\,\,        0 \,\,  & \,\,   0 \\
0\,\, &\,\,        1\,\,  & \,\,   0  \\
0\,\, &\,\,        0\,\,   &\,\,   1
\end{array}
\right),
\label{yr}
\end{equation}
where $h(s_W)=(1-4s^2_W)^{1/2}$.
Notice that the right-handed neutral currents remain diagonal, but 
it is not so for the left-handed ones. Here the matrices ${\cal O}^u_L$ and 
${\cal O}^d_L$ survive in a different form than they appear in the 
$V_{CKM}$ matrix. These matrices ${\cal O}^u_L$, ${\cal O}^d_L$ and 
$V_{CKM}$ have to be only determined from experiment
since we cannot set none of them to be diagonal in a natural way.

\subsection{Lepton sector}
\label{subsec:edml}
Beside the neutron EDM, another ones which are experimentally well studied  
(alas not definitively yet) is the EDM of the electron and the muon.
It is interesting that the EDM of the electron ($d_e$) in the standard model is 
rather small, of the order of magnitude of $2\times 10^{-38}$ $e$ cm~\cite{hoo}. 
However, the value of $d_e$ could be even smaller, of the order of $10^{-41}$ 
$e$ cm if there is a cancellation of the three-loop diagrams~\cite{kh} even if 
QCD corrections are included~\cite{bo}.
On the other hand, the experimental upper limit is  $\leq 4\times 10^{-27}$ $e$ 
cm~\cite{ele}. Hence, a measurement of a nonzero large
electron EDM (and other elementary particles like muon, tau and neutron, see 
below) would indicate new physics beyond the standard model.

Although we have $CP$ violation in the singly charged scalar sector,
there is no contribution to the EDM of the leptons at the 1-loop level since 
neutrino are massless. If neutrinos remain
massless the contribution to electric dipole moment will arise at the 
three-loop level~\cite{chao}.

This is not the case for diagrams
involving doubly charged scalars and the known leptons and the 
exotic ones (see Fig.~6). It is always possible to choose the doubly charged
scalars those fields which carry this phase {\it i.e.}, the $CP$ violation 
in the leptonic sector occurs only through the exchange of both exotic leptons 
and of doubly charged scalar fields. 

The Yukawa couplings of leptons with the doubly charged scalars are
given in Eq.~(\ref{dcyul1}) where the scalar fields are still symmetry 
eigenstates. In fact, there is one Goldstone boson $G^{++}$ and a physical 
one $Y^{++}$; we denote its mass by $m_{Y^{++}}$. 
(In this model there is not lepton number violation
in the interactions with the neutral scalars.)
As we have shown  in Secs..~\ref{subsec:leptons} and \ref{subsec:gl} it is not 
possible to absorb all phases in the complete lepton lagrangian density.
Since neutrinos are considered massless here, the only contributions to the 
leptonic EDM arise from the doubly charged scalars as shown in Fig.~6.
From this we obtain
\begin{equation}
d_l=-\frac{em_l}{32\pi^2m^2_{Y^{++}}}
\,\sqrt{2}M^2_WG_F\;{\bf O}_{l}\;
\sin(2\theta_\alpha),\quad l=e,\mu,\tau;
\label{edml}
\end{equation}
where we have defined
\begin{equation}
{\bf O}_{l}=\sum_j{\cal K}^2_{lj}\,
\;\frac{4m^2_{E_j}}{M^2_U}\;[F_+(m_l,m_{E_j})+F_-(m_l,m_{E_j})],
\label{oo}
\end{equation}
the matrix ${\cal K}$ has been introduced in Eqs.~(\ref{hclm1}) and 
(\ref{dcyul1}); 
\begin{equation}
F_\pm(m_l,m_{E_j})\!\!=\!\!-\frac{m^2_{Y^{++}}}{2m_l^2}\ln\frac{m^2_{Y^{++}}}
{m^2_{E_j}}+
\frac{m^2_{Y^{++}}}{2m^2_l\Delta_l} \left(m^2_{Y^{++}}\pm m^2_l-m^2_{E_j}\right)
\,\ln\left[ \frac{m^2_{E_j}+m^2_{Y^{++}}-m^2_l+\Delta_l}
{m^2_{E_j}+m^2_{Y^{++}}-m^2_l-\Delta_l}\right],
\label{fs}
\end{equation}
and
\begin{equation}
\Delta^2_l=(m^2_{Y^{++}}+m^2_{E_i}-m^2_l)(m^2_{Y^{++}}-m^2_{E_i}-m^2_l).
\label{delta2}
\end{equation}

For nondegenerate heavy leptons the mixing angles remain in Eq.~(\ref{oo}).
For instance, the contribution of $E_1$ to the electron EDM, using $m_{E_1}=50$ 
GeV and $m_{Y^{++}}=100$ GeV~\cite{pdg} is given by
\begin{equation}
d_e\approx 10^{-17}\,\left(\frac{M^2_W}{M^2_U}\right)\,
{\cal K}^2_{e1}\,
\sin(2\theta_\alpha)\;e\,{\rm  \, cm}.
\label{edme}
\end{equation}
Assuming $M_U=300$ GeV and the factor with the mixing angles 
${\cal K}^2_{e1}\approx10^{-8}$ we obtain $d_e\approx 10^{-27}\,e$ cm for any 
value of $\theta_\alpha$, which is compatible with the experimental upper limit 
of $10^{-27}\,e$ cm~\cite{ele}.

For the muon we have an experimental EDM's upper limits
of~$<10^{-19}\,e$ cm~\cite{muon}. It means a constraint in 
${\cal K}^2_{\mu2}\leq 1$.
For the tau lepton a limit of $10^{-17}\,e$ cm is derived from 
$\Gamma(Z\to \tau^+\tau^-)$~\cite{escribano1,escribano2}.
In the present model it is at least as large as 
$10^{-19}\,e$ cm. In particular, notice that from Eq.~(\ref{edml})
\begin{equation}
\frac{d_\mu}{d_e}= \frac{m_\mu\,\sum_j
{\cal K}^2_{\mu j}
m^2_{E_j}F(m_\mu,m_{E_j})}
{m_e\,\sum_j{\cal K}^2_{e j}
m^2_{E_j}F(m_e,m_{E_j})}.
\label{edm2}
\end{equation}
Using $F_+(m_l,m_{E_j})=F_-(m_l,m_{E_j})\equiv F(m_l,m_{E_j})
\approx -(m^2_Y/2m^2_l)\ln(m^2_Y/m^2_{E_j})$
we obtain from Eq.~(\ref{edm2}) 
\begin{equation}
\frac{d_\mu}{d_e}=\frac{m_e}{m_\mu}\,
\frac{\sum_j{\cal K}^2_{\mu j}m^2_{E_j}
\ln(m^2_X/m^2_{E_j})}
{\sum_j{\cal K}^2_{e j}m^2_{E_j}
\ln(m^2_X/m^2_{E_j})}.
\label{edm3}
\end{equation}

Notice that if
\begin{equation}
\frac{\sum_j{\cal K}^2_{\mu j}m^2_{E_j}
\ln(m^2_X/m^2_{E_j})}
{\sum_j{\cal K}^2_{e j}m^2_{E_j}
\ln(m^2_X/m^2_{E_j})}\gg \frac{m_\mu}{m_e}.
\label{edm4}
\end{equation}
is satisfied, we can have $d_\mu\gg d_e$. For instance,  
assuming that $E_1$($E_2$) dominates the EDM of the electron (muon) this 
condition implies that (neglecting the logarithmic in both the numerator 
and denominator) 
\begin{equation}
\left\vert \frac{{\cal K}_{\mu 2}m_{E_2}}
{{\cal K}_{e 1}m_{E_1}}\right\vert\approx
10^{4}\,\sqrt{\frac{m_\mu}{m_e}}\sim10^{5},
\label{edm5}
\end{equation}
this value of the ratio $\vert{\cal K}_{\mu 2}m_{E_2}
/{\cal K}_{e 1}m_{E_1}\vert$ is easily obtained for the ${\cal K}_{e1}$ and
${\cal K}_{\mu2}$ given above if $m_{E_2}\approx 10m_{E_1}$. 
We see that in fact, the EDM of the muon can be larger than the EDM of the 
electron. A similar situation happens with the tau lepton if $E_3$ dominates 
here,
\begin{equation}
\left\vert\frac{{\cal K}_{\tau 3}m_{E_3}}
{{\cal K}_{e 1}m_{E_1}}\right\vert
\approx 10^{5}\,\sqrt{\frac{m_\tau}{m_e}}\sim6\times 10^{6}.
\label{edm6}
\end{equation}
If Eqs.~(\ref{edm5}) and (\ref{edm6}) are satisfied, the
muon and the lepton tau can have an EDM which is as large as their respective 
present experimental limit. For example, analyzing the process 
$e^+e^-\to \tau\tau\gamma$, the L3 collaboration has obtained the value
$d_\tau=(0.0\pm1.5\pm1.3)\times 10^{-16}\,e$ cm~\cite{l398}. 
The conditions in Eqs.(\ref{edm5}) and (\ref{edm6}) are equivalent to the
assumption that the mixing matrix ${\cal K}\equiv
\left({\cal O}^e_L\right)^T {\cal O}^E_L$ is almost completely diagonal.
If the tau lepton has in fact a large EDM it will induce anomalous couplings
of the $Z$ boson to fermions which could be be seen in tau-charm 
factories~\cite{tao}.

In this model there is no rare decays such as $\mu\to 3e$ at tree level. 
However, the same loop diagrams that contribute for the EDM of the leptons
imply transition magnetic and electric moments, like $\mu\to e\gamma$.
However, it will constrain only the matrix elements 
$\left[\left({\cal O}^e_L\right)^T {\cal O}^E_L\right]^2_{\mu1}$.

\section{Conclusions}
\label{sec:con}
The model of $CP$ violation of this work seems like an admixture of both
spontaneous breaking as in Lee's two doublets model~\cite{tdlee} and the 
charged-Higg-boson exchange of Weinberg's three doublets one~\cite{sw1}. 
However, some differences are important to be pointed
out. In a pure charged-scalar-exchange where the phases are in the coupling
constants of the scalar potential it was shown that there is also $CP$ 
violation through the exchange of neutral Higgs 
bosons~\cite{dema}. This is correct in models with $SU(2)_L\otimes U(1)_Y$ 
symmetry with several scalar doublets, 
that is, with all of them having the same quantum number. For this reason a more
general mixing among the scalar fields of the same charge is possible.
In the present model all triplets have different $U(1)_N$ charge so this 
constraint their interaction terms. Hence, there is not mixing among the real
and imaginary part of the neutral scalar fields even with three scalar 
triplets as can be seen from Eqs.~(\ref{nfm}) and (\ref{nsxm}). 
If we had considered $a_{10}\not=0$ in the
scalar potential in Eq.~(\ref{potential}) we would have $CP$ violation in the
propagator of single charged scalar like in the Weinberg model but since the 
$a_{10}$ term in the scalar potential does not contribute to the mass matrix of 
the neutral Higgs there is not mixing among $CP$ even and $CP$ odd neutral 
scalars. 

There are other interesting features of this model: {\it i)}
Notice that since we have no $CP$ violation in the neutral scalar sector the
Weinberg's three-gluon operators involving neutral Higgs boson exchanges do not 
contribute to the EDM of the neutron at all~\cite{sw2}. There is also no 
contribution to the EDM of the electron and neutron coming from the diagrams of 
Ref.~\cite{bzee} which involve $CP$ violation in the propagator of the neutral 
Higgs bosons and it does not depend on the phase convention since in 
Eqs.~(\ref{yuunf}) and (\ref{e6}) phases do not appear at all;
{\it ii)} In this model radiative corrections up to 2-loop level will not induce
phases in the CKM matrix. These type of diagrams
are the same which would contribute to the penguin diagram, say
of the $\epsilon'_K$ parameter, hence there is no
penguin contributions to the $CP$ violating parameters at least at the 2-loop 
level. Hence, it means that $arg det M$ vanish up to this loop order.
Furthermore, if we assume that $CP$ is also conserved in the pure QCD part 
then $\theta=0$ at the tree level and $\bar{\theta}$ will be finite and 
calculable. Of course, this is not such a natural solution to the $\theta$-vacua
problem but at least it is in the same foot than the assumption that all Yukawa 
couplings are real at the tree level.

This model of $CP$ violation seems like a particular realization of the 
soft superweak model proposed recently by Georgi and Glashow~\cite{gg}. In that
model the violation of $CP$ is due to
the coupling of the left-handed doublet to the heavy sector in the mass
eigenstate basis involve a complex matrix. In our model, the heavy sector
corresponds to the exotic quarks $J,j_n$, the exotic leptons $E_j$ and the
scalars $Y^+$ and $Y^{++}$; the only complex numbers are the phase 
$\theta_\alpha$ appearing
in the trilinear term in the scalar potential and the complex VEVs (only 
$\theta_\chi$ after using the $SU(3)$ freedom to eliminate two phases). 

With three triplets and one sextet which are needed in the model of 
Ref.~\cite{ppf} it is possible to have truly spontaneous violation of the $CP$ 
symmetry~\cite{laplata1}. 
In this case, the minimization condition of the scalar potential 
implies more complicated constraint equations on the imaginary part
of the neutral scalars so that two of the phases of the VEV survive in the
lagrangian density. The phenomenology of this model has been studied 
in Ref.~\cite{laplata2}.

\acknowledgments 
This work was supported by Funda\c{c}\~ao de Amparo \`a Pesquisa
do Estado de S\~ao Paulo (FAPESP), Conselho Nacional de 
Ci\^encia e Tecnologia (CNPq) and by Programa de Apoio a
N\'ucleos de Excel\^encia (PRONEX). One of us (VP) gratefully acknowledge
useful discussions with C. O. Escobar.

\newpage

\begin{center}
{\bf Figure Captions}
\end{center}
\vskip .5cm

\noindent {\bf Fig.~1} Tree level contributions to $\Delta M_K$.\\

\noindent {\bf Fig.~2} Some of the box diagram contributions to 
$\varepsilon$ and $M_{12}$.\\ 
  
\noindent {\bf Fig.~3} The box diagram contribution to 
$\varepsilon'$.\\ 
 
\noindent {\bf Fig.~4} Possible diagram contributing to the EDM of the quarks 
$d$.\\

\noindent {\bf Fig.~5} Possible diagram contributing to the EDM of the quarks 
$u$.\\

\noindent {\bf Fig.~6} Diagrams contributing to the EDM of the charged leptons.


\begin{figure*}
\mbox{\epsfxsize=430pt \epsffile{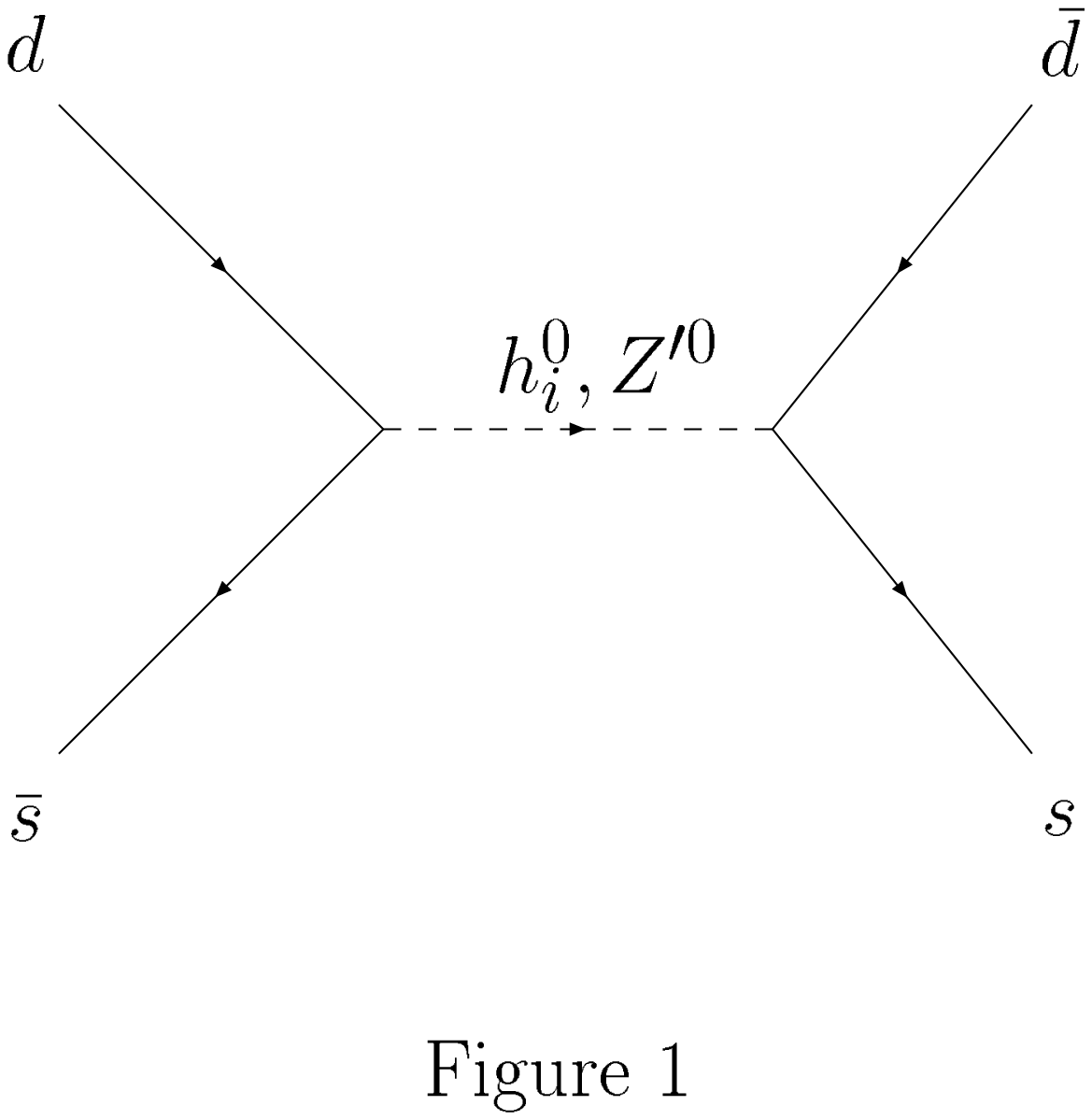}}
\end{figure*}

\begin{figure*}
\parbox[c]{6.0in}{\mbox{\qquad\epsfig{file=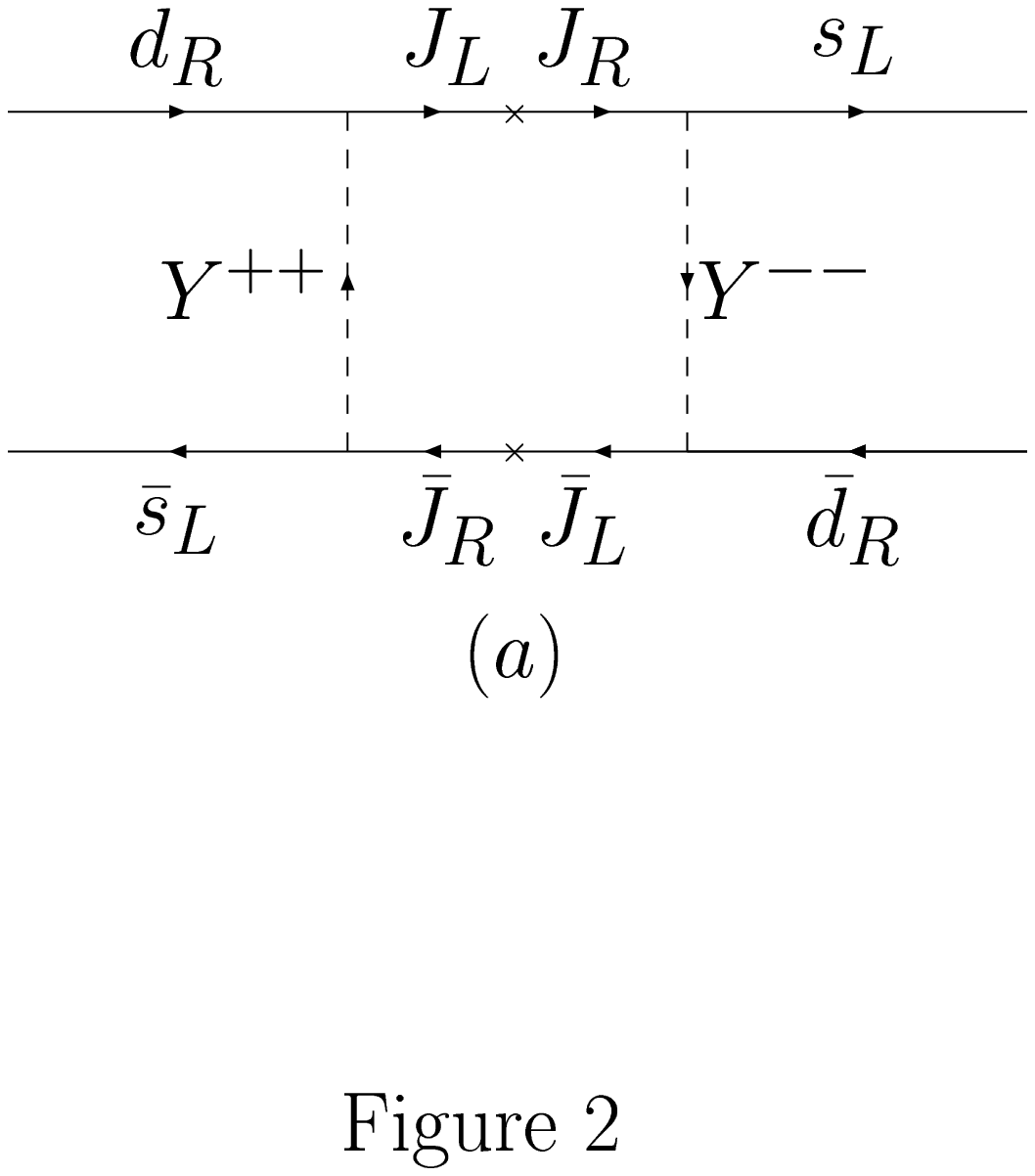,width=\linewidth}}}
\end{figure*}

\begin{figure*}
\mbox{\epsfxsize=430pt \epsffile{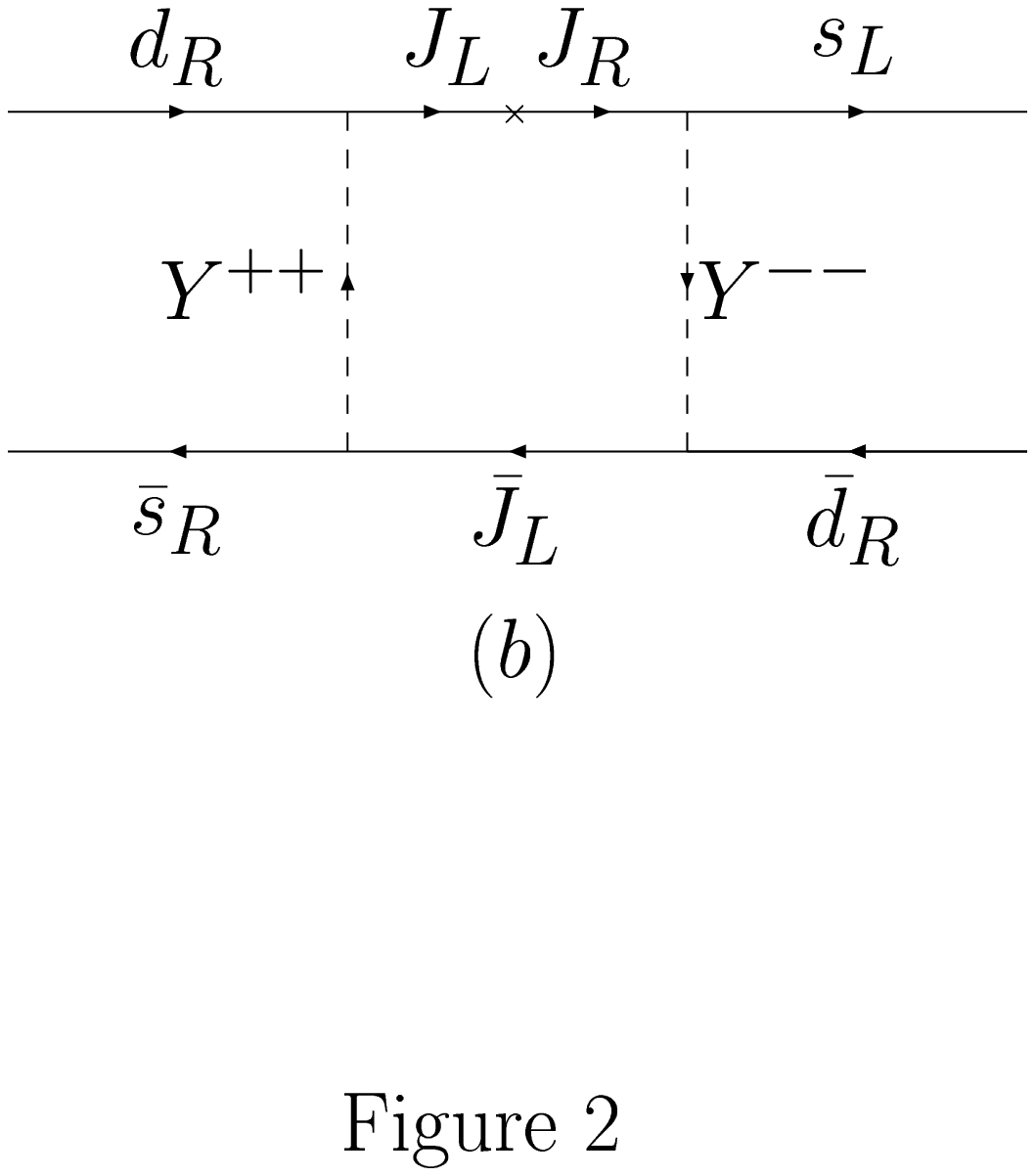}}
\end{figure*}

\begin{figure*}
\mbox{\epsfxsize=430pt \epsffile{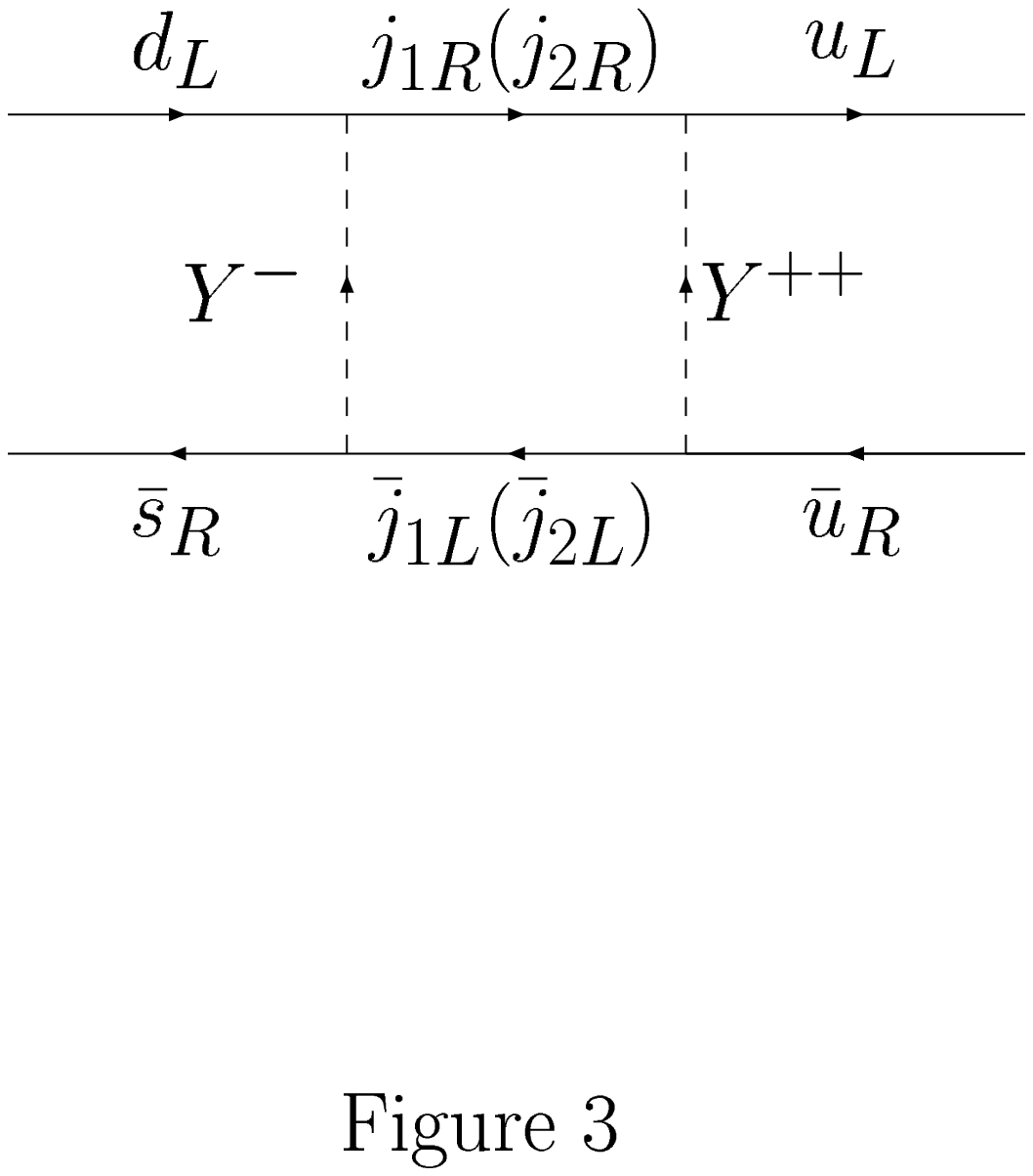}}
\end{figure*}

\begin{figure*}
\mbox{\epsfxsize=430pt \epsffile{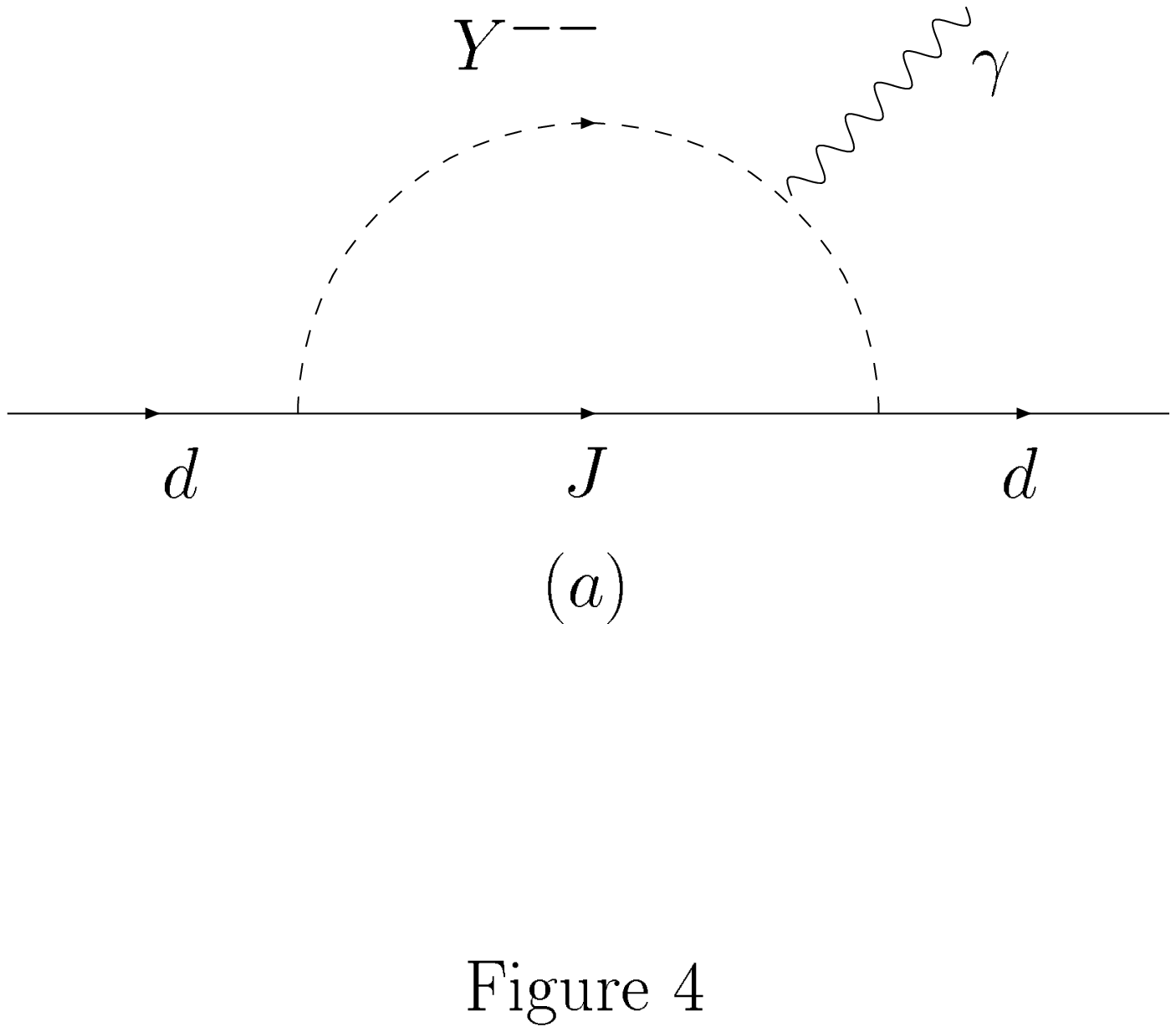}}
\end{figure*}

\begin{figure*}
\mbox{\epsfxsize=430pt \epsffile{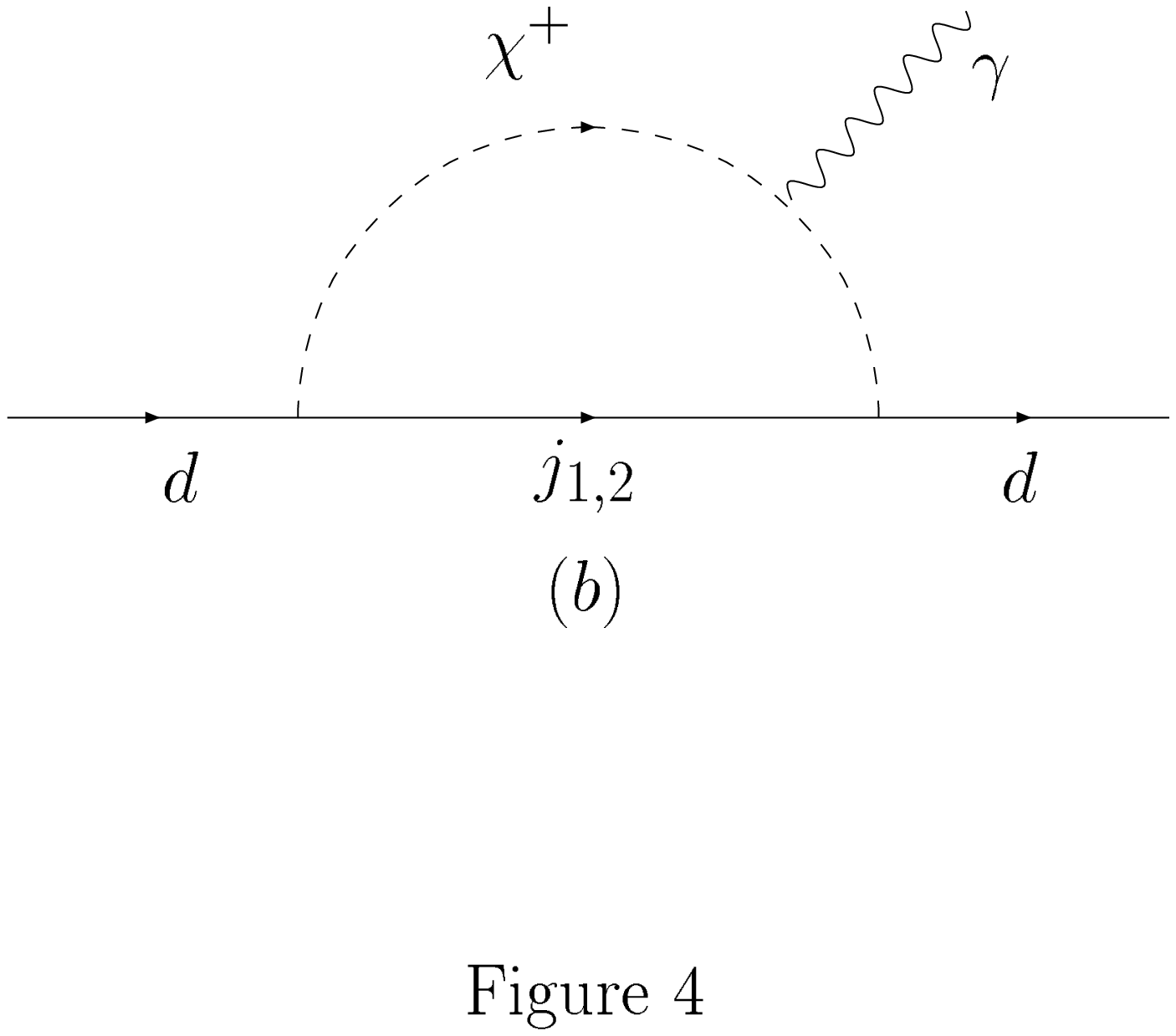}}
\end{figure*}

\begin{figure*}
\mbox{\epsfxsize=430pt \epsffile{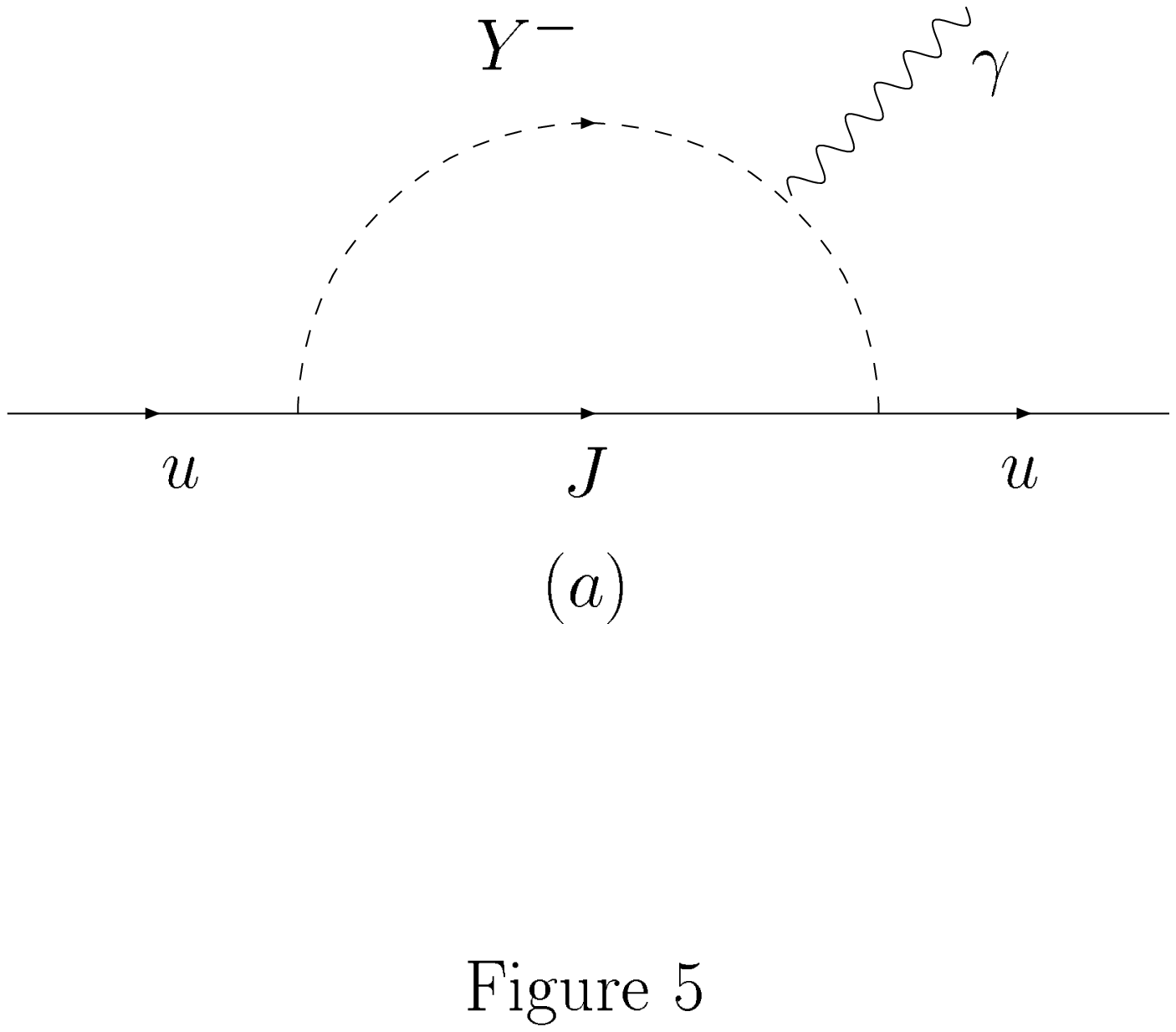}}
\end{figure*}

\begin{figure*}
\mbox{\epsfxsize=430pt \epsffile{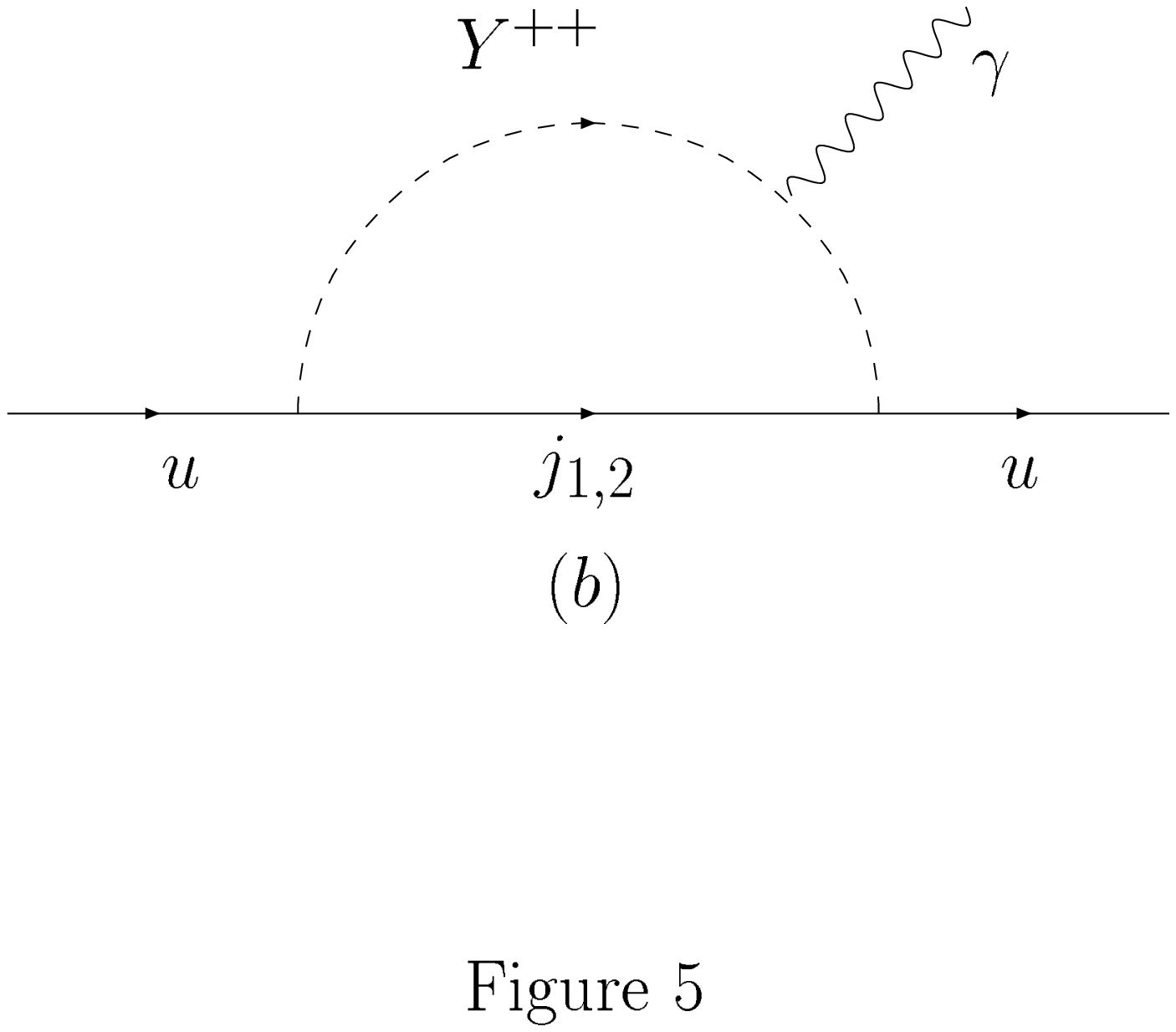}}
\end{figure*}

\begin{figure*}
\mbox{\epsfxsize=430pt \epsffile{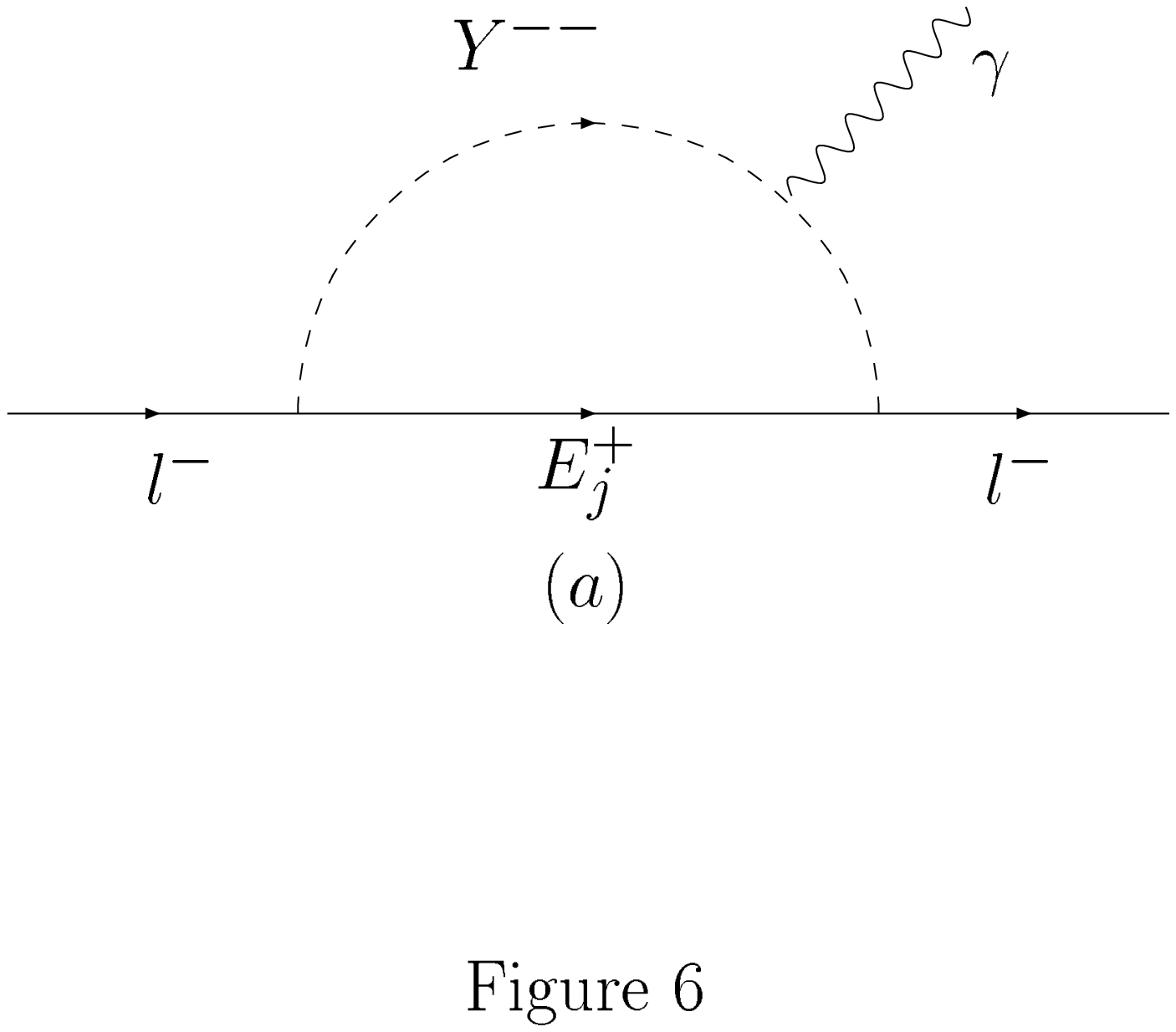}}
\end{figure*}

\begin{figure*}
\mbox{\epsfxsize=430pt \epsffile{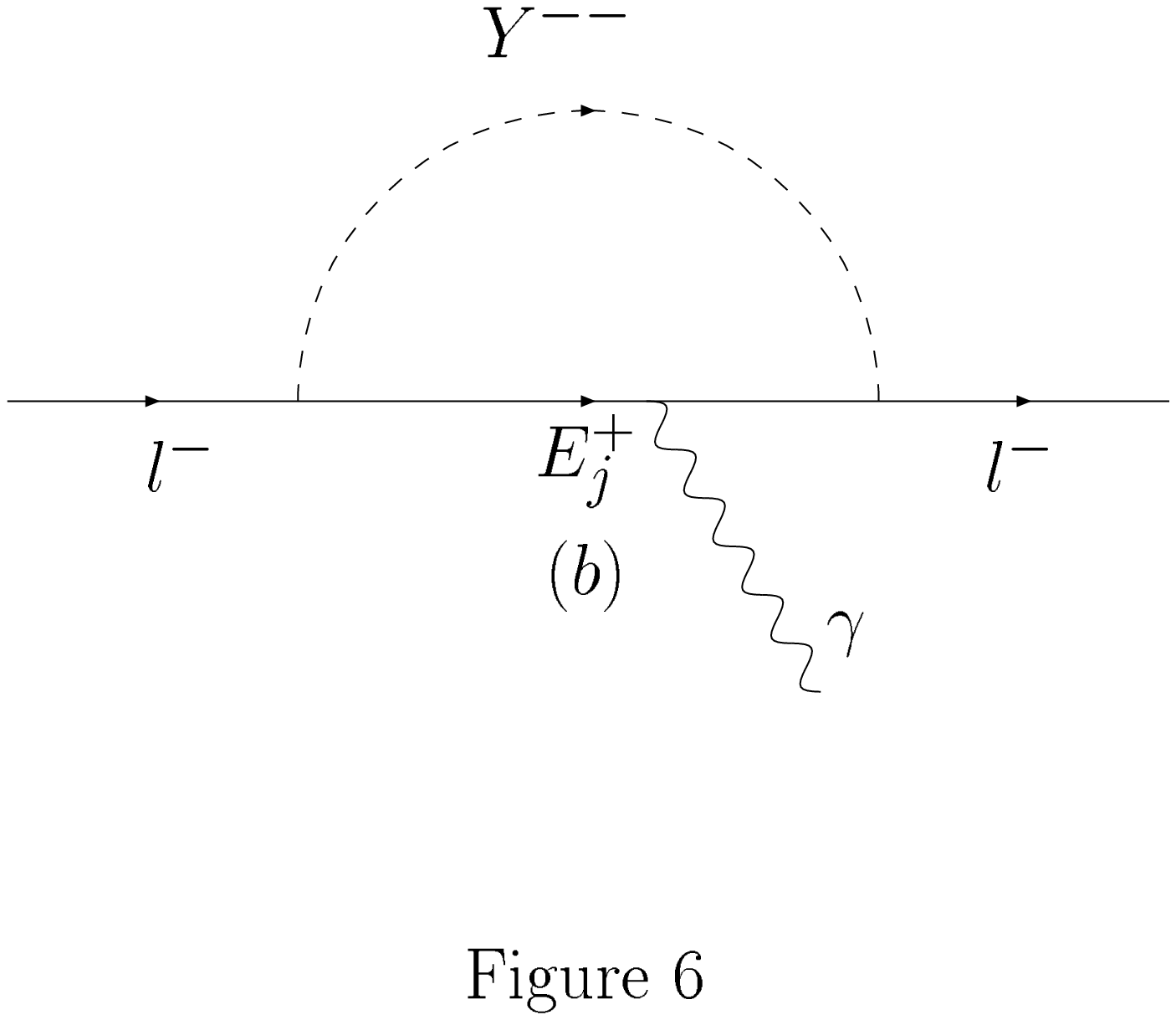}}
\end{figure*}
\end{document}